\documentstyle[12pt,amsbsy,scr]{article}   
\makeatletter
\@ifundefined{scr}{\newcommand{\scr}{\cal}}{\relax}
\makeatother
\newcommand{\bs}{\boldsymbol}
\newcommand{\fd}{\!\!\!/}

\newcommand{\mrm}{\mathrm}  

\begin{document}
\input{psfig}
\begin{center}

{\bf MESON PHOTO- AND ELECTROPRODUCTION}
\renewcommand{\thefootnote}{\fnsymbol{footnote}}\footnote[4]{
This work was supported by the Deutsche Forschungsgemeinschaft (SFB201)}
\vspace{5mm}

{\sl Lothar Tiator}\\
\vspace{5mm}

{\small
Institut f\"ur Kernphysik, Universit\"at Mainz, 55099 Mainz, Germany}
\end{center}
\vspace{5mm}

\begin{abstract}
Photo- and electroproduction of pions and eta-mesons is studied on the
nucleon. The formalism of such reactions is presented including
polarization degrees of freedom and theoretical approaches are discussed.
In particular, the threshold and $\Delta$-resonance region of pion
photoproduction is considered and recent developments on low energy
theorems (LET) and the E2/C2 quadrupole excitation of the $\Delta$ are
presented. For eta photoproduction we show our calculations for the
threshold region of the new Mainz experiment and demonstrate the
sensitivity of this reaction to the elementary $\eta NN$ coupling and the
resonance parameters of the $S_{11}(1535)$. In nuclear applications we
point out the importance and advantage of light nuclei as $d$, $^3H$, $^3He$
and $^4He$ for the isospin structure of the reactions as well as for medium
effects of the nucleon resonances.
\end{abstract}
\vspace{5mm}

\begin{center}
{\bf 1. Introduction}
\end{center}

Besides the nucleon, the pion is the most important particle in nuclear
physics. It is the mediator of nuclear forces and is responsible for the
long-range part of nucleon-nucleon interaction. Similar to the neutron,
the pion may also exist as a stable particle in the nuclear environment, due
to binding effects the particle is far off-shell and cannot decay. Its
presence, however, can be observed in scattering experiments, e.g. as meson
exchange currents in electron scattering. In nuclear reactions, where the $
cm$ energy is large enough, the pions can be produced on-shell in processes
such as NN scattering, nuclear collisions and photo- or electroproduction on
nucleons and nuclei. The general properties of pions and etas are listed in
table 1.

The theory of pion photoproduction was written in the 1950s. Kroll and
Ruderman \cite{Kro54} were the first to derive model-independent
predictions in the threshold region, so-called low energy theorems (LET),
by applying gauge and Lorentz invariance to the reaction $\gamma + N
\rightarrow \pi + N$. The general formalism for this process was developed
by Chew et al \cite{Che57} (CGLN amplitudes). Fubini et al \cite{Fub65}
extended the earlier predictions of LET by including also the hypothesis of
a partially conserved axial current (PCAC). This way they succeeded in
describing the threshold amplitudes as a power series in the ratio
$\mu=m_\pi/m_N$ up to terms of order $\mu^2$. For energies up to 500 MeV
dispersion theoretical calculations \cite{Ber67} and data analyses
\cite{Ber71,Pfe72,Cra83,Arn90} exist and results are tabulated in terms of
the leading multipoles.

\begin{table}[htbp]
\begin{center}
\renewcommand{\arraystretch}{1.2}    
\begin{tabular}{c|cccccc}\hline
 &       &        & Mass    & Lifetime & Decay modes  &  \\
 & $I^G$ & $J^{PC}$ & (MeV) & (sec) & & (\%) \\ \hline
 $\pi^\pm$ & $1^-$ & $0^-$ & $139.57$ & $2.60\cdot 10^{-8}$ &
             $\mu\nu_\mu$  & (100)\\
 $\pi^0$ & $1^-$ & $0^{-+}$ & $134.97$ & $8.4\cdot 10^{-17}$ &
             $\gamma\gamma$  & (98.8)\\
 & & & & &   $\gamma e^+ e^-$  & (1.2)\\
 $\eta$ & $0^+$ & $0^{-+}$ & $547.45$ & $5.5\cdot 10^{-19}$ &
             $\gamma\gamma$  & (38.8)\\
 & & & & &   $3\pi^0$        & (31.9)\\
 & & & & &   $\pi^+\pi^-\pi^0$ & (23.6)\\
\hline
\end{tabular}
\end{center}
\begin{center}
{\small {\bf Table 1:}
Basic properties of $\pi$ and $\eta$ mesons}
\end{center}
\end{table}

In 1969 Peccei \cite{Pec69} wrote down an effective chiral Lagrangian for
single-pion photoproduction that included a phenomenological
$\pi N \Delta$ interaction. After that such models have been very
successful in the description of pion production up to 500 MeV. The main
difference in the various approaches is the treatment of the $\Delta$
excitation, the modelling of the mass and the energy-dependent width.
Blomqvist and Laget \cite{Blo77} first gave a non-relativistic operator
expansion up to order $(p/m)^2$ which could be easily incorporated in
nuclear physics applications \cite{Tia80,Tia84}. Although they reproduced
the elementary cross sections on the nucleon very well, the original
version violated unitarity which became especially apparent in selected
channels of coherent $(\gamma,\pi^0)$ from nuclei \cite{Koc84} and in the
reaction $^{14}N(\gamma,\pi^+)^{14}C_{g.s.}$ \cite{Tia88}.
However, unitarity can be
enforced and modern versions of pion photo- and electroproduction operators
are unitarized in the most important $\Delta$ channel \cite{Lag88} or as in
the work of Davidson et al \cite{Dav91} in all partial waves. Another
interesting approach is the dynamical model of Nozawa, Blankleider and Lee
\cite{Noz90}. Treating the $\pi N$ scattering and pion photoproduction
consistently in a coupled-channels calculation this model satisfies
unitarity by construction. However, in its present form it is based on
purely phenomenological $\pi N$ separable potentials.

\begin{figure}[htbp]
\centerline{\parbox{9cm}{
\vspace{8cm} }}
\hfill \parbox{6.5cm}{\small{\bf \mbox{Fig. 1:}}
The total photoabsorption cross section of the proton and its decomposition in
a few selected channels.}
\end{figure}

Experimentally, pion photoproduction has been extensively studied in the 1960s
and 1970s at Bonn, DESY, Lebedev and many others places. A compilation of
the data
was done in 1978 by Menze et al \cite{Men77} and Fujii et al \cite{Fuj77}.
For pion electroproduction
most of the data has been taken at Bonn, DESY, Frascati and  NINA,
an overview can
be found in the books of Amaldi, Fubini and Furlan \cite{Ama79}.
Towards the end of the 80s, after completion of the first modern electron
accelerators with high duty factor new experimental activity has started
with higher resolution and higher accuracy. It then came as a big surprise
when in $\gamma,\pi^0$ experiments on the proton at Saclay \cite{Maz86} and
Mainz \cite{Bec90} the threshold $E_{0+}$ amplitude was much smaller than
predicted by low energy theorems (LET). However after careful analyses the
experimental data are finally in agreement with LET at the $(\gamma,\pi^0)$
threshold point. What remains to be explained is a strong energy variation
of the s-wave amplitude over a range of 10 MeV above threshold. Again a
new Mainz experiment that is currently being analysed is expected to
clarify this important question.

Single-pion production above the threshold region is governed by the
excitation of nucleon resonances. Below the two-pion threshold it is
practically the only source that contributes to the total photoabsorption
cross section, shown in fig. 1. The M1 excitation of the first, the E1 of the
second and the E2 excitation of the third resonance region show up as
significant peaks; the multi-pion background appears rather flat in
the inclusive data. More than one order of magnitude below appears the
contribution of the eta production which, however, turns out to be a very
clean signal of a particular nucleon resonance, the $S_{11}(1535)$.

\begin{figure}[htbp]
\vspace{8cm}
{\small {\bf Fig. 2:}
Total cross sections for $\pi^0$ and $\pi^+$ photoproduction on the proton.
The data are from the compilations of Fujii et al \cite{Fuj77} and Menze
et al \cite{Men77}.}
\end{figure}

Fig. 2 shows the total cross sections for $\pi^0$ and $\pi^+$ production on
the proton, the sum of both yields the total photoabsorption (hatched area
in fig. 1) up to the $E2$ excitations. While the $\pi^0$ reaction is almost
entirely given by the $\Delta(1232)$ resonance, the $\pi^+$ reaction is
also very sensitive to $E1$ excitations, the pion-nucleon dipole moment at
threshold and the electric dipole resonances around 1.5 GeV $cm$ energy.
The latter can be understood from the isospin decomposition, charged pion
production is most sensitive to $I=1/2 (N^*)$ resonances, while neutral
pion production as a probe favour $I=3/2 (\Delta)$ resonances.

In section 2, we describe the general formalism for coincidence cross
sections of $(e,e'\pi)$ and $(e,e'\eta)$ on nucleons or other spin 1/2 targets.
In particular, we discuss the response functions that include
polarization degrees of
freedom for photon, the target and the recoil nucleon. The invariants of meson
electroproduction in the CGLN basis and the multipole decomposition are
given in section 3. In section 4 we discuss models of pion production, in
particular the field theoretical effective Lagrangian method. Furthermore,
the role of nucleon resonances is outlined. In sections 5 and 6 we present
data and calculations of pion photoproduction at threshold and in the
$\Delta$-resonance region. The eta photoproduction is derived in section 7
and consequences for the $\eta NN$ coupling and the $S_{11}(1535)$
resonance parameters are shown. In nuclear
applications we discuss pion and eta photoproduction on light nuclei in
section 8. Finally we give a summary with an outlook into future
experimental work.

\begin{center}
{\bf 2. Coincidence cross section}
\end{center}

Following the notation of Bjorken and Drell \cite{Bjo64}, the differential
cross section for an exclusive process may be written
\begin{eqnarray} \label{elec1}
 d\sigma & = & \frac{\varepsilon _{i}}{k_{i}}\frac{m_{e}}{\varepsilon _{i}}
\frac{m_{i}}{E_{i}}\frac{m_{e}}{\varepsilon _{f}}\frac{d^{3}k_{f}}{(2\pi )
^{3}}\frac{1}{2\omega _{\pi }}\frac{d^{3}k}{(2\pi )^{3}}\frac{m_{f}}{E_{f}}
\frac{d^{3}P_{f}}{(2\pi )^{3}}(2\pi )^{4}\delta^{(4)}(P_{i}+q-k-P_{f})
                                                       \nonumber\\[0.2cm]
& & \times \mid \langle k_{f} \mid j_{\mu } \mid k_{i} \rangle
 q^{-2} \langle P_{f},k\mid J^{\mu }\mid P_{i}\rangle
     \mid ^{2},
\end{eqnarray}
where the phase space is evaluated in the laboratory frame and all
kinematical variables are defined as in fig. 3.

\begin{figure}[htbp]
\vspace{6cm}
{\small {\bf Fig. 3:}
The kinematic variables for: {\bf a)} pion
photoproduction and {\bf b)} a two-arm electroproduction experiment in the
one-photon exchange approximation. }
\end{figure}

We have assumed a purely electromagnetic process described by the currents
of the electron ($j_\mu$) and the hadronic system ($J_\mu$). The current of
the electron is well defined in terms of the Dirac spinors with
normalization $\bar u u = 1$ and the Dirac matrices $\gamma_\mu$.

The square of the invariant matrix element in eq. (\ref{elec1}) can be
written as a product of 2 rank-2 Lorentz tensors $\eta_{\mu \nu}$ and
$W_{\mu \nu}$. The leptonic tensor $\eta_{\mu \nu}$ is well defined from QED
and
for initially polarized electrons with helicity $h$ it is evaluated as
\newpage
\begin{eqnarray} \label{elec2}
\eta _{\mu \nu} & = & \sum_{s_{f}}(\bar u(k_{f},s_{f})e\gamma _{\mu}
u(k_{i},s_{i}))(\bar u(k_{f},s_{f})e\gamma _{\nu }u(k_{i},s_{i}))^{\ast }
\nonumber\\
& = & \frac{e^{2}}{2m_{e}^{2}}(2K_{\mu }K_{\nu }+\frac{1}{2}q^{2}g_{\mu \nu}
      -\frac{1}{2}q_{\mu}q_{\nu}+ih\varepsilon _{\mu \nu \alpha \beta}
      q^{\alpha}K^{\beta})
\end{eqnarray}
with $m_e$ the electron mass, $K=\frac{1}{2}(k_i + k_f)$, $q=k_i - k_f$ and
$\varepsilon_{\mu\nu\alpha\beta}$ a completely antisymmetrized tensor with
$\varepsilon_{0123}=1$. The hadronic tensor is defined as
\begin{equation} \label{elec3}
W_{\mu \nu} = (\frac{m_{N}}{4\pi W})^{2}\langle \chi_{f}\mid J_{\mu}\mid
\chi_{i}\rangle \langle \chi_{f}\mid J_{\nu}\mid \chi_{i} \rangle ^{\ast},
\end{equation}
where $J^\mu = (\rho, \bs J)$ is the hadronic current operator for a meson
electroproduction process and will be specified in the next chapter.

Fig. 4 shows the kinematics in a typical coincidence experiment with the
scattering plane ($x,z$) of the electron and the reaction (or production)
plane of the outgoing pion. The coordinates are defined as
\begin{equation} \label{elec4}
\hat{\bs e}_{x} = \hat{\bs e}_{y}\times
\hat{\bs e}_{z},\quad
\hat{\bs e}_{y} = \frac{\hat{\bs k}_{i}\times
\hat{\bs k}_{f}}{\sin \Theta _{e}},\quad
\hat{\bs e}_{z} = \hat{\bs q}
\end{equation}
with the scattering angle $\Theta_e$ of the electron in the lab frame.

\begin{figure}[hbtp]
\vspace{6cm}
{\small {\bf Fig. 4:}
Kinematics of a typical coincidence experiment leading to the out-of-plane
production of a pion.}
\end{figure}

In an electroproduction experiment the polarization of the virtual photon
is controlled by the electron kinematics, the transverse and longitudinal
polarizations
\begin{equation} \label{elec5}
\varepsilon = (1+\frac{2{\boldsymbol q}^{2}}{Q^{2}}\tan ^{2}
\frac{\Theta_{e}}{2})^{-1}\quad \text{and}\quad
\varepsilon_{\text{L}} = \frac{Q^{2}}{\omega^{2}}\varepsilon,
\end{equation}
where all variables are evaluated in the lab frame. In an electron
scattering experiment the four-momentum transfer of the virtual photon $q^2$
is always negative (space-like) and it is common to define the positive
quantity $Q^2 = -q^2$.

Combining eqs. (\ref{elec1}-\ref{elec5}) we can write the coincidence cross
section in the following form \cite{Ama79,Don78}
\begin{eqnarray} \label{elec6}
\frac{d\sigma}{d\Omega_{f}d\varepsilon_{f}d\Omega_{\pi}} & = & \Gamma
\frac{d\sigma_{v}}{d\Omega_{\pi}}\\
\frac{d\sigma_{v}}{d\Omega_{\pi}} & = &
\frac{d\sigma_{T}}{d\Omega_{\pi}}
+\varepsilon_{\text{L}}\frac{d\sigma_{L}}{d\Omega_{\pi}}
+[2\varepsilon_{\text{L}}(1+\varepsilon)]^{\frac{1}{2}}
\frac{d\sigma_{TL}}{d\Omega_{\pi}}\cos\Phi_\pi\\
& & +\varepsilon\frac{d\sigma_{TT}}{d\Omega_{\pi}}\cos 2\Phi_\pi
+h[2\varepsilon_{\text{L}}(1-\varepsilon)]^{\frac{1}{2}}
\frac{d\sigma_{TL'}}{d\Omega_{\pi}}\sin\Phi_\pi
+h(1-\varepsilon^{2})^{\frac{1}{2}}\frac{d\sigma_{TT'}}{d\Omega_{\pi}},
\nonumber
\end{eqnarray}
where
\begin{equation}
\Gamma = \frac{\alpha}{2\pi ^{2}}\frac{\varepsilon_{f}}{\varepsilon_{i}}
\frac{k_{\gamma}}{Q^{2}}\frac{1}{1-\varepsilon}
\end{equation}
is the flux of the virtual photon field. In this expression we have
introduced the `photon equivalent energy', $k_\gamma = (W^2-m_i^2)/2m_i$,
the laboratory energy necessary for a real photon to excite a hadronic
system with the $cm$ energy $W$. The first two terms of eq. (7), the
transverse (T) and longitudinal (L) cross sections do not depend on the
azimuthal angle. The third term and the fifth term describe transverse-
longitudinal interferences (TL and TL'), due to their dependence on
$\cos \Phi_\pi$ and sin $\Phi_\pi$ they have to contain an explicit factor
$\sin \Theta_\pi$, i.e. they vanish along the axis of momentum transfer. The
same is true for the fourth term, a transverse-transverse interference (TT)
proportional to $\sin^2 \Theta_\pi$. The last term (TT') can only be
observed by target or recoil polarization.

The cross section of eq. (7) can be expressed in response functions as
\begin{eqnarray} \label{elec9}
\frac{d\sigma_\alpha}{d\Omega}=\frac{k}{k_\gamma^{cm}}R_\alpha,\quad
\alpha=\{\text{ L, T, TL, TT, TL', TT'}\},\quad
k_\gamma^{cm}=\frac{m_i}{W} k_\gamma,
\end{eqnarray}
which are linear combinations of matrix elements of the hadronic tensor
$W_{\mu\nu}$:

\parbox{5cm}{
\begin{eqnarray*}
R_{T} & = &  \frac{1}{2}(W_{xx}+W_{yy}),\nonumber\\
\cos \Phi_{\pi}R_{TL} & = & -Re W_{xz},\nonumber\\
\cos 2\Phi_{\pi}R_{TT} & = & \frac{1}{2}(W_{xx}-W_{yy}),\nonumber
\end{eqnarray*}}
\hfill \parbox{9cm}{
\begin{eqnarray} \label{elec10}
R_{L} & = &  W_{zz},\nonumber\\
\sin \Phi_{\pi}R_{TL'} & = & Im W_{yz},\\
R_{TT'} & = & Im W_{xy}.\nonumber
\end{eqnarray}}

For unpolarized electrons ($h$=0) the number of response functions reduces
to four and in case of forward or backward production ($\theta_\pi=0\;
\text{or}\;\pi$) or for an inclusive experiment, where the mesons are not
observed, only the longitudinal and transverse response functions remain
and can be separated by a Rosenbluth plot like in single-arm electron
scattering experiments. A separation of the LT interference term is also
very simple by measuring left and right of the virtual photon axis in the
scattering plane ($\Phi_\pi=0\;\text{or}\;\pi$). Finally, in order to
separate also the TT structure function an out-of-plane experiment is
necessary to disentangle the longitudinal and the transverse/transverse
contributions.

However, even in an out-of-plane experiment with full separation of the
structure functions of eq. (7) the hadronic current $J_\mu$ cannot be
fully determined. As we will see later the current can be expressed in 6
invariant complex amplitudes requiring 11 linear independent measurements
for a complete experiment. These additional measurements are possible in
experiments with polarization degrees of freedom of the target and recoil
nucleons.

The general formalism for coincidence and polarization experiments has been
developed by Donnelly and Raskin \cite{Don86,Ras89}. Recently, this
formalism has been applied to the reaction $p(\vec{e},e'\vec{p})\pi^0$,
in order to study the structure of nucleon resonances \cite{Lou90,Dre92}.
Due to
the polarization of either the target nucleon or the recoiling nucleon the
differential cross section contains a total of 18 structure functions,
\begin{eqnarray} \label{elec11}
\frac{d\sigma_{v}}{d\Omega} & = & \frac{k}{k_{\gamma}^{\text{CM}}}
\{(R_{T}+P_{n}R_{T}^{n})+\epsilon_{\text{L}}(R_{L}+P_{n}R_{L}^{n}) \nonumber\\
& & +[2\epsilon_{\text{L}}(1+\epsilon)]^{\frac{1}{2}}
[(R_{TL}+P_{n}R_{TL}^{n})\cos
\Phi+(P_{l}R_{TL}^{l}+P_{t}R_{TL}^{t})\sin\Phi] \nonumber\\[0.2cm]
& & +\epsilon[(R_{TT}+P_{n}R_{TT}^{n})\cos 2\Phi+(P_{l}R_{TT}^{l}+
P_{t}R_{TT}^{t})\sin 2\Phi] \nonumber\\[0.1cm]
& & +h[2\epsilon_{\text{L}}(1-\epsilon)]^{\frac{1}{2}}
[(R_{TL'}+P_{n}R_{TL'}^{n})
\sin\Phi+(P_{l}R_{TL'}^{l}+P_{t}R_{TL'}^{t})\cos\Phi]   \nonumber\\[0.1cm]
& & +h(1-\epsilon^{2})^{\frac{1}{2}}(P_{l}R_{TT'}^{l}+P_{t}R_{TT'}^{t})\},
\end{eqnarray}
where $n,l,t$ can be both $n_i,l_i,t_i$ for target polarization or
$n_f,l_f,t_f$ for recoil polarization. Even more structure functions are
possible for target-recoil asymmetries, where in addition to the polarized
target also the polarization of the recoiling nucleon has to be measured.
Such experiments, however, are beyond the purpose of the present paper.

In addition to the electron helicity $h$, in eq. (11) there appear the
projections of the proton spin unto the three axes $\hat{\bs n}=\hat{\bs q}
\times\hat{\bs k}/\sin\theta_{\pi}$ (normal to the reaction plane),
$\hat{\bs l}$ (along the proton momentum) and $\hat{\bs t}=\hat{\bs n}
\times\hat{\bs l}$. For example, $P_n=\hat{\bs n}\cdot\hat{\bs S}_R$ is the
projection of the spin vector (in the proton rest frame!) unto the axis
normal to the reaction plane. As shown in fig. 5 our coordinate system $\{t,n
,l\}$ for measuring the polarizations is only one particular choice. Other
choices are e.g. the system $\{x,y,z\}$ often used for target polarization
and the system $\{x',y',z'\}$ for recoil polarization \cite{Fas92}. In the
review of Drechsel and Tiator \cite{Dre92} we have given explicit formulas
and multipole decompositions for all of the 18 structure functions in the
system $\{t,n,l\}$. These can be transformed to the other coordinate
systems by simple rotations. For target polarization in $\{x,y,z\}$:
\begin{eqnarray} \label{elec12}
R_\alpha (x) & = & -\cos \theta R_\alpha(t_i) - \sin \theta R_\alpha(l_i)
\nonumber\\
R_\alpha (y) & = & R_\alpha (n_i) \nonumber\\
R_\alpha (z) & = & \sin \theta R_\alpha(t_i) - \cos \theta R_\alpha(l_i)
\end{eqnarray}
and for recoil polarization in $\{x',y',z'\}$:
\begin{equation} \label{elec13}
R_\alpha (x') = -R_\alpha (t_f),\quad R_\alpha (y') = R_\alpha (n_f),\quad
R_\alpha (z') = -R_\alpha (l_f).
\end{equation}

\begin{figure}[htbp]
\vspace{8cm}
{\small {\bf Fig. 5:}
Orientation of the different coordinate systems $\{x,y,z\}$, $\{x',y',z'\}$
and $\{t,n,l\}$  used for target and recoil polarization. Note that
$\hat{\bs y}=\hat{\bs y}'=\hat{\bs n}$ and is pointing upwards out of the
production plane.}
\end{figure}

The cross section for reactions induced by real photons follows from eq.
(\ref{elec11}) if we (i) replace $k_\gamma^{cm}$ by the energy of the real
photon in the $cm$ frame and (ii) drop all longitudinal currents. We find
\begin{eqnarray} \label{elec14}
\frac{d\sigma}{d\Omega} & = & \frac{k}{q}
\{(R_{T}+P_{n}R_{T}^{n})\\
& & +\Pi_{T}[(R_{TT}+P_{n}R_{TT}^{n})\cos 2\varphi
-(P_{l}R_{TT}^{l}+P_{t}R_{T}^{t})\sin 2\varphi]\nonumber\\[0.1cm]
& & +\Pi_{C}(P_{l}R_{TT'}^{l}+P_{t}R_{TT'}^{t})\},\nonumber
\end{eqnarray}
where $\Pi_{T}$ is the degree of linear polarization of the photon and
$\varphi$ the angle of the polarization vector relative to the reaction
plane, and $\Pi_{C}$ is the degree of circular polarization. For example,
for completely linearly polarized photons and polarization vector normal to
the production plane, $\Pi_T$=1, $\varphi$=$\pi$/2 and $\Pi_C$=0, while
right (left)-circularly polarized photons have $\Pi_T$=0 and $\Pi_C$=+1(-1).
Note that in an experiment with linearly polarized real photons the
out-of-plane angle $\varphi$ is measured from the production plane towards the
photon polarization vector and is therefore opposite to the azimuthal angle
$\Phi_\pi$ of the pion in an electroproduction experiment.

In experiments with polarization degrees of freedom it is common to define
the following observables \cite{Fas92,Moo78,Wal69}:
\begin{description}
\item[(i)] the polarized photon asymmetry (beam asymmetry)
\begin{equation}
\Sigma(\theta) = -R_{TT}/R_{T}
\end{equation}
\item[(ii)] the polarized target asymmetry
\begin{equation}
T(\theta) = R_T(n_i)/R_{T} = -R_{TT}(n_f)/R_{T}
\end{equation}
\item[(iii)] the recoil polarization
\begin{equation}
P(\theta) = R_T(n_f)/R_{T} = -R_{TT}(n_i)/R_{T},
\end{equation}
\end{description}
where $n_i$ and $n_f$ refers to the polarization of the nucleon (nucleus)
in the initial and final state and $\theta$ is the pion angle in the c.m.
system.

In addition there are four observables for polarization of beam and target
($E,F$, $G,H$), four of beam and recoil ($O_x,O_z,C_x,C_z$)
and four of target and recoil polarization ($T_x,T_z,L_x,L_z$)
\cite{Bar75,Fas92}.
As shown by Barker et al \cite{Bar75} a complete experiment requires,
besides the differential cross section and the three single polarization
observables five double polarization observables, provided that no four of
them come from the same set (beam-target, beam-recoil and target-recoil).

Alternatively we can define the single polarization observables in the
following way
\begin{eqnarray}
\Sigma & = & \frac{d\sigma/d\Omega^{\perp} -
d\sigma/d\Omega^{\parallel}}{d\sigma/
d\Omega^{\perp} + d\sigma/d\Omega^{\parallel}}\\
T & = & \frac{d\sigma/d\Omega^{(+)} -
d\sigma/d\Omega^{(-)}}{d\sigma/d\Omega^{(+)}
+ d\sigma/d\Omega^{(-)}}\\
P & = & \frac{d\sigma/d\Omega^{(+)} -
d\sigma/d\Omega^{(-)}}{d\sigma/d\Omega^{(+)}
+ d\sigma/d\Omega^{(-)}}
\end{eqnarray}
where $+(-)$ refers to the initial or final nucleus polarized parallel
(antiparallel) to the $\hat{\bs n}$-axis, and $\perp (\parallel)$ refers to a
photon
linearly polarized perpendicular (parallel) to the production plane.

\newpage
\begin{center}
{\bf 3. Invariants and multipoles}
\end{center}

The current operator appearing in the hadronic tensor of eq.~(\ref{elec3})
has the normal symmetry behaviour of an electromagnetic current, i.e.
scalar charge $\rho$ and vector current $\bs J$, as long as the pion field
$\bs \Phi$ is included. However, if we drop the pion wave function which
carries an intrinsic parity $0^-$, the structure of the current in Dirac or
Pauli space is pseudoscalar for $\rho$ and pseudovector for $\bs J$. The
most general form can be given by the CGLN amplitudes $F_i$ in Pauli space
\cite{Che57}
\begin{eqnarray} \label{inv1}
{\bs J} & = & \frac{4\pi W}{m_{N}}\left[i\tilde{\bs\sigma}F_{1}
+({\bs\sigma}\cdot\hat{\bs k})({\bs\sigma}\times\hat{\bs q})F_{2}
+i\tilde {\bs k}({\bs\sigma}\cdot\hat{\bs q})F_{3}
\right.\nonumber\\
& & \left.+i\tilde {\bs k}({\bs\sigma}\cdot \hat {\bs k})F_{4}
+i\hat {\bs q}({\bs\sigma}\cdot\hat{\bs q})F_{5}
+i\hat {\bs q}({\bs\sigma}\cdot\hat{\bs k})F_{6}\right],\nonumber\\
\rho & = & \frac{4\pi W}{m_{N}}\left[i({\bs\sigma}\cdot
\hat {\bs k})F_{7}
+i({\bs\sigma}\cdot \hat {\bs q})F_{8}\right] =
\frac{{\bs q}\cdot {\bs J}}{\omega }
\end{eqnarray}
with the unit vectors of photon (pion) in the $cm$ frame $\hat{\bs q}$
($\hat{\bs k}$) and $\tilde{\bs \sigma}=\bs \sigma-(\bs \sigma \cdot
\hat{\bs q}) \hat{\bs q}$ etc. In order to be consistent with previous
notations for the CGLN amplitudes and their multipole decompositions, we
have introduced a factor $4\pi W/m_N$. In this way the current can be
calculated directly from Feynman diagrams in the notation of Bjorken and
Drell. The structure functions $F_1$, $F_2$, $F_3$ and $F_4$ describe the
transverse current while the longitudinal component is given by $F_5$ and
$F_6$. $F_7$ and $F_8$ are the amplitudes of the charge and are related to
the longitudinal ones by current conservation, $k_\mu J^\mu = 0$:
\begin{equation} \label{inv2}
F_7 = \frac{q}{\omega} F_6 \quad \text{and} \quad
F_8 = \frac{q}{\omega} F_5.
\end{equation}
These structure functions depend on three variables, e.g. the square of the
4-momentum transfer $Q^2$ and on two of the Mandelstam variables (s and t
or, alternatively, $\omega_L$ and $\theta_\pi^{cm}$). Due to the strong
interaction in the $\pi N$ system, they have complex values. Therefore,
there are six absolute values and five relative phases that have to be
determined in each kinematical situation. Since the number of independent
structure functions is connected with the spin degrees of freedom of the
interacting particles, a complete determination of the structure functions
requires polarization experiments as mentioned before. Instead of the CGLN
amplitudes one often uses helicity amplitudes defined by transitions
between eigenstates of the helicities of nucleon and photon, see e.g.
\cite{Ama79}.

For a further analysis of photo- and electroproduction experiments it is
useful to decompose the CGLN amplitudes $F_i$ into multipoles. In
electroproduction spin and angular momentum have to be distinguished
between initial and final states. In the initial state the photon carries
spin 1 and has orbital angular momentum $l_\gamma$ relative to the target
nucleon. Its wave functions can be characterized by vector spherical
harmonics,
\begin{equation}
{\bs Y}_{l_{\gamma}LM}=\sum_{\nu}^{}C(1\lambda,l_{\gamma}\nu\mid LM)
\hat{\bs e}_{\lambda}Y_{l_{\gamma}\nu}(\hat{\bs \Omega}).
\end{equation}
The transverse polarizations $\lambda=\pm1$ leading to electric and
magnetic multipole transitions $EL$ and $ML$, the longitudinal polarization
$\lambda=0$ leading to longitudinal or Coulomb transitions $CL$.

The final state is described by an orbital momentum $l$ of the pion
relative to the recoiling nucleon with parity $(-1)^{l+1}$ due to the
intrinsic parity of the pion. The total spin of the final state, $J$, has
to equal the total spin of the initial state,
\begin{equation}
J=\mid l\pm\frac{1}{2}\mid = \mid L\pm\frac{1}{2}\mid.
\end{equation}
Using parity arguments, we find for
\begin{eqnarray}
CL, EL: && (-1)^L = (-1)^{l+1} \rightarrow \mid L-l\mid = 1\nonumber\\
    ML: && (-1)^{L+1} = (-1)^{l+1} \rightarrow L = l.
\end{eqnarray}
In photoproduction the multipoles are classified according to the final
state with a capital letter to indicate the electromagnetic nature,
$E_{l\pm}, M_{l\pm}$, where the $\pm$ sign is an abreviation for
$J=l\pm\frac{1}{2}$ and has nothing to do with parity. In electroproduction
both longitudinal $L_{l\pm}$ and scalar (time-like) $S_{l\pm}$ are defined,
related by $\omega S_{l\pm} = q L_{l\pm}$ due to current conservation
(gauge invariance).

As an example, the lowest electromagnetic excitation modes and the
corresponding states of the pion-nucleon system are given in table 2. The
first two columns denote the familiar electromagnetic multipoles. In the
next two columns we find the spin and angular momentum of the $\pi N$
system. As an example, the $\Delta(1232)$ resonance with $J=\frac{3}{2}$
and $l=1$ can be excited by $M1, E2$ and $C2$ radiation. Its multipoles are
denoted by $M_{1+}, E_{1+}$ and $L_{1+}$, respectively.
\renewcommand{\arraystretch}{1.2}             
\begin{table}
\begin{center}
\begin{tabular}{clccl}
\hline
 & Electromagnetic & $\pi N$ & system & Pion production \\
$L$ & multipole       & $J$     & $l$    & multipole \\
\hline
0 & $C0$    & 1/2 & 1 & $L_{1-}$ \\[3mm]
1 & $E1, C1$ & 1/2 & 0 & $E_{0+}, L_{0+}$\\
  &         & 3/2 & 2 & $E_{2-}, L_{2-}$\\
  & $M1$    & 1/2 & 1 & $M_{1-}$\\
  &         & 3/2 & 1 & $M_{1+}$\\[3mm]
2 & $E2, C2$ & 3/2 & 1 & $E_{1+}, L_{1+}$\\
  &         & 5/2 & 3 & $E_{3-}, L_{3-}$\\
  & $M2$    & 3/2 & 2 & $M_{2-}$\\
  &         & 5/2 & 2 & $M_{2+}$\\ \hline
\end{tabular}
\end{center}
\begin{center}
{\small {\bf Table 2:}
Amplitudes for pion electroproduction}
\end{center}
\end{table}

The structure functions can be decomposed into a multipole series in terms
of derivatives of the Legendre polynomials $P_l$
\begin{eqnarray} \label{cgln}
F_{1} & = & \sum_{l\geq0}\{(lM_{l+}+E_{l+})P_{l+1}^{\prime}
+[(l+1)M_{l-}+E_{l-}]P_{l-1}^{\prime}\}, \nonumber\\
F_{2} & = & \sum_{l\geq1}[(l+1)M_{l+}+lM_{l-}]P_{l}^{\prime}, \nonumber\\
F_{3} & = & \sum_{l\geq1}[(E_{l+}-M_{l+})P_{l+1}^{\prime\prime}
+(E_{l-}+M_{l-})P_{l-1}^{\prime\prime}],   \nonumber\\
F_{4} & = & \sum_{l\geq2}(M_{l+}-E_{l+}-M_{l-}-E_{l-})P_{l}^{\prime\prime},
\nonumber\\
F_{5} & = & \sum_{l\geq0}[(l+1)L_{l+}P_{l+1}^{\prime}
-lL_{l-}P_{l-1}^{\prime}],  \nonumber\\
F_{6} & = & \sum_{l\geq1}[lL_{l-}-(l+1)L_{l+}]P_{l}^{\prime}.
\end{eqnarray}
The Legendre polynomials are functions of the polar angle of the pion in
the $cm$ frame, $\theta=\theta_{\pi}^{cm}$. The electric and magnetic
multipoles, $E_{l\pm}$, $M_{l\pm}$ and the longitudinal multipoles $L_{l\pm
}$ depend on both energy $W$ and momentum transfer $Q^2$. For
photoproduction, $Q^2=0$, and $E_{l\pm}$, $M_{l\pm}$ depend only on the
energy. The longitudinal multipoles $L_{l\pm}$ cannot be measured with
real photons but are finite in the limit of $Q^2\rightarrow 0$. In the
region up to about 500 MeV, the leading multipoles have been analysed
or calculated in dispersion theory and tabulated by \cite{Ber67, Ber75,
Pfe72, Cra83, Arn90}.

Finally, a further decomposition of the multipoles in isospin space is
necessary in order to relate $\gamma,\pi$ and $\pi N$ scattering.
The initial state is characterized by the target nucleon with isospin
$\frac{1}{2}$ coupling to the electromagnetic current with isoscalar ($1$)
and isovector ($\tau_0$) components. In the final state, the pion is an
isovector particle ($\Phi^\alpha$) coupling to the nucleon with
($\tau_\alpha$). The Pauli matrices appearing in the interaction of the nucleon
with the photon and pion may be arranged in the overall matrix element in a
symmetrical form
\begin{equation} \label{iso}
A = \frac{1}{2}A^{(-)}[\tau_{\alpha},\tau_{0}]+A^{(+)}\delta_{\alpha0}
+A^{(0)}\tau_{\alpha},
\end{equation}
where the first two terms are the commutator and anticommutator of
$\tau_\alpha$ with the isovector electromagnetic current while the last term
corresponds to the isoscalar current. In terms of the three isospin
amplitudes, the four physical amplitudes are obtained as

\parbox{5cm}{
\begin{eqnarray*}
A(\gamma p \rightarrow n\pi^+) & = & \sqrt{2}(A^{(0)}+A^{(-)})\\
A(\gamma n \rightarrow p\pi^-) & = & \sqrt{2}(A^{(0)}-A^{(-)})
\end{eqnarray*}}
\hfill \parbox{8cm}{
\begin{eqnarray}
A(\gamma p \rightarrow p\pi^0) & = & A^{(+)}+A^{(0)}\nonumber\\
A(\gamma n \rightarrow n\pi^0) & = & A^{(+)}-A^{(0)}.
\end{eqnarray}}

If the $\pi N$ final state and in particular nucleon resonances are
analysed in terms of isospin $I$, the three amplitudes of eq. (27) have to
be combined as

\parbox{6cm}{
\begin{eqnarray*}
&& A^{(3/2)}=A^{(+)}-A^{(-)}\\
&& A^{(1/2)}=A^{(+)}+2A^{(-)}\\
&& A^{(0)}
\end{eqnarray*}}
\hfill \parbox{8cm}{
\begin{eqnarray}
&&(I=3/2)^{\vphantom{(3/2)}} \hspace{5cm} \nonumber\\
&&(I=1/2)^{\vphantom{(3/2)}} \nonumber\\
&&(I=1/2)^{\vphantom{(0)}}
\end{eqnarray}}\\
or vice versa
\begin{eqnarray}
A^{(+)}& = & \frac{1}{3}(2A^{(3/2)}+A^{(1/2)})\nonumber\\
A^{(-)}& = & \frac{1}{3}(A^{(1/2)}-A^{(1/2)}).
\end{eqnarray}
As an example, the dominant multipoles of pion photoproduction, the
electric dipole amplitude for charged pion photoproduction $Re E_{0+}(\pi^+
n)$ (mainly $Re E_{0+}^{(-)}$) and the magnetic dipole amplitude for neutral
pion photoproduction, $Im M_{1+}(\pi^0p)$ (mainly $M_{1+}^{(3/2)}$), are
shown in fig. 6.

\begin{figure}[htbp]
\vspace{8cm}
{\small {\bf Fig. 6:}
The dominant multipoles of pion photoproduction. The data are from Pfeil
and Schwela \cite{Pfe72} (circles), and Berends and Donnachie \cite{Ber75}
(triangles).}
\end{figure}

Since the electromagnetic interaction is much weaker than the $\pi N$ final
state interaction, the phases of the $\gamma,\pi$ multipoles are determined
by the $\pi N$ phase shifts. From unitarity of the S-matrix and
time-reversal invariance the 2-channel ($\pi N, \gamma N$) S-matrix can be
written as
\begin{eqnarray}
S_{\alpha} = \left( \begin{array}{c@{\quad\quad}cc}
\eta e^{2 i \delta_\alpha} & i \sqrt{1- \eta^2}
e^{i (\delta_\alpha + \delta_\beta)}\\
\vspace{4pt}
i \sqrt{1- \eta^2} e^{i (\delta_\alpha + \delta_\beta)}
& \eta e^{2 i \delta_\beta}
\end{array} \right) = S_{\alpha}^{\dagger},
\end{eqnarray}
where $\alpha = \{\ell, J, I\}$, $\delta_\alpha$ is the $\pi N$ scattering
phase shift and $\eta$ is a real number. The small Compton scattering phase
$\delta_\beta$ can be neglected in practical calculations, therefore we get
the Watson-Theorem \cite{Wat54}, which relates the phase of pion production
multipoles directly to the $\pi N$ scattering phases, up to a minus sign,
\begin{eqnarray}
A_\alpha^{(I)} = \left| A_\alpha^{(I)} \right|
e^{i \delta_{\alpha} + i n \pi}.
\end{eqnarray}
As an example, for the $\Delta$ excitation multipoles we find
\begin{eqnarray}
M_{1+}^{(3/2)} & = & \left| M_{1+}^{(3/2)} \right|
e^{i \delta_{33}}\nonumber\\
E_{1+}^{(3/2)} & = & - \left| E_{1+}^{(3/2)} \right|
e^{i \delta_{33}}
\end{eqnarray}
In model calculations, e.g. effective Lagrangians with nucleon resonance
excitations, very often unitarity is not a priori satisfied because of the
neglect of higher order $\pi N$ rescattering. In these cases Watson's
theorem can be enforced by introducing a background phase $\delta_B$ and
a correction phase $\Phi$ with \cite{Dav91}
\begin{eqnarray}
A_\alpha^{(I)} & = & A_\alpha^{(I)}\text{(Born)}\cdot e^{i\delta_B}
+ A_\alpha^{(I)}\text{(Resonance)}\cdot e^{i\Phi}\nonumber\\
& = & \pm \mid A_\alpha^{(I)}\mid\cdot e^{i\delta_\alpha}.
\end{eqnarray}
As a further approximation, the small background phase is often neglected.

\begin{figure}[htbp]
\vspace{9cm}
{\small {\bf Fig. 7:}
Pion-nucleon phase shifts $\delta_\alpha$ as functions of the kinetic energy in
the lab frame. The curves are parametrizatons of Nozawa et al \cite{Noz90}.}
\end{figure}

In fig. 7 we show the leading s- and p-wave phase shifts for $\pi N$
scattering together with a fit from the dynamical model by Nozawa et al
\cite{Noz90}. In the region below 500 MeV, practically the only relevant
channel, where unitarity plays an important role is the $P_{33}$ with the
$\delta_{33}$ phase going through $90^\circ$ at the resonance position of
the $\Delta(1232)$. For this channel we also show from ref. \cite{Noz90}
the electric and magnetic multipoles (fig. 8) which reflect the unitarity
condition discussed before.

\begin{figure}[htbp]
\vspace{9cm}
{\small {\bf Fig. 8:}
Real and imaginary parts of the $M_{1+}$ and $E_{1+}$ multipoles in the 3/2
channel. The figure and curves are from Nozawa et al \cite{Noz90} and
show their dynamical model. The dashed lines are obtained when the $\Delta$
term in the production operator is set to zero.}
\end{figure}

\newpage
\begin{center}
{\bf 4. Models for pion production}
\end{center}

On the phenomenological basis, the first theory of nucleons and pions was
formulated with a pseudoscalar (PS) coupling. Such a theory can be
renormalized. However, it does not obey the PCAC relation. The axial
current is not conserved in the soft-pion limit.

The Lagrangian is given in terms of the free Lagrangians of a nucleon
obeying the Dirac equation, a pion following the Klein-Gordan equation
and a PS interaction term
\begin{eqnarray} \label{mod1}
\scr L & = & \bar\psi(i\partial\fd-m)\psi + \frac{1}{2}(\partial^\mu\bs\Phi
\cdot\partial_\mu\bs\Phi-m_\pi\bs\Phi^2) + \scr L_{int}^{PS},\\
& & \scr L_{int}^{PS}=-ig_{\pi NN}\bar\psi\gamma_{5}{\bs \tau}\cdot\psi{\bs
\pi}.
\end{eqnarray}
Using minimal coupling ($\partial_\mu\rightarrow\partial_\mu+i e A_\mu$)
a conserved electromagnetic current can be constructed
\begin{eqnarray} \label{mod3}
J_\mu^{em} & = & J_\mu^{N} + J_\mu^{\pi}\nonumber\\
& = & e \bar\psi\gamma_\mu\frac{1+\tau_3}{2}\psi +
e (\bs\Phi\times\partial_\mu \bs\Phi)_3.
\end{eqnarray}
However, the axial current is not conserved and in the limit of zero pion
mass, the divergence gives $\partial^\mu J_{5\mu}^\alpha = -g_{\pi NN}
\bar\psi \psi
\Phi^\alpha$. This problem can be solved in an alternative model using
pseudovector (PV) coupling of pions and nucleons,
\begin{equation} \label{mod4}
\scr L_{int}^{PV}=\frac{f_{\pi NN}}{m_{\pi}}\bar\psi\gamma_{\mu}
\gamma_{5}{\bs \tau}\cdot\partial^{\mu}{\bs \Phi}\psi
\end{equation}
{}From the equivalence theorem, i.e. PS and PV coupling result in the same
strength for one-pion exchange in NN interaction,
\begin{equation} \label{mod5}
\frac{f_{\pi NN}}{m_\pi}=\frac{g_{\pi NN}}{2 m}=\frac{g_A}{2 f_\pi}
\qquad \text{with}
\qquad \frac{g_{\pi NN}^2}{4 \pi}=14.
\end{equation}
The axial current of the PV model yields
\begin{equation} \label{mod6}
J_{5\mu}^\alpha=\bar\psi \gamma_\mu \gamma_5 \frac{1}{2} \tau_\alpha \psi
+ f_\pi \partial_\mu \Phi^\alpha
\end{equation}
fulfilling the PCAC relation
\begin{equation} \label{mod7}
\partial^\mu J_{5\mu}^\alpha = - f_\pi m_\pi^2 \Phi^\alpha.
\end{equation}

Due to the derivative $\partial^\mu$ in the PV interaction Lagrangian of
eq.~(\ref{mod4}), an additional e.m. current arises from minimal coupling
\begin{equation} \label{mod8}
J_{\mu}^{\pi N}=-i \frac{e f_{\pi NN}}{m_\pi}\varepsilon_{\alpha\beta 3}
\bar\psi \gamma_\mu \gamma_5 \tau_\alpha \psi \Phi^\beta
\end{equation}
The e.m. current of the nucleon in eq. (\ref{mod3}) contains only the
Dirac current $\gamma_\mu$ leading to the normal magnetic moment of a Dirac
particle. In a renormalizable pion-nucleon theory, the anomalous magnetic
moments of proton and neutron arise from higher order pion loops (vertex
corrections). In an effective theory, however, we work at the tree-level
and the anomalous magnetic moments are introduced as a Pauli current
\begin{equation} \label{mod9}
J_\mu^{Pauli}=\frac{e}{2 m}\partial^\nu \bar\psi(\kappa_S+\kappa_V\tau_0)
\sigma_{\mu\nu}\psi
\end{equation}
with the isoscalar $\kappa_S=(\kappa_p+\kappa_n)/2=-0.06$ and isovector
$\kappa_V=(\kappa_p-\kappa_n)/2=+1.85$ anomalous moments of the nucleon.

\begin{figure}[htbp]
\vspace{6cm}
{\small {\bf Fig. 9:}
Diagrams contributing to pion photoproduction: (a) direct, (b) crossed
nucleon pole, (c) pion pole, (d) contact or seagull, (e) and (f) isobar
excitation, (g) triangle anomaly (vector meson exchange), (h) square
anomaly, (i) rescattering term.}
\end{figure}

Together with the PV interaction (\ref{mod4}) and the e.m. currents
(\ref{mod3},\ref{mod8},\ref{mod9}) all ingredients for pion photoproduction
are given and the
tree-level Feynman diagramms of order $e f_{\pi NN}$ can be
calculated (fig. 9a-d).
The first two are the nucleon s- and u-channel pole terms and are
calculated with the Dirac and Pauli current and the PV interaction. The
third diagram is the pion pole term from the e.m. current of the pion
and the PV interaction. This and the last Born term (fig. 9d)
which is the seagull or contact term from eq.~(\ref{mod8}) occur only for
charged pion production. Therefore in $\pi^0$ production and also in $\eta$
production only the nucleon s- and u-channel diagrams exist. If we
calculate the Born terms in the alternative PS model, the seagull term that
was constructed from the PV Lagrangian by minimal coupling does not
exist. Due to the equivalence theorem, the pion-pole term does not change,
however, the nucleon-pole terms with an off-shell nucleon in the
intermediate state produce quite different results. Part of this difference
simply produces the same result as the seagull term in the PV model, this
is the consequence of the underlying low energy theorem from gauge invariance
which is fulfilled in both models. The next to leading order terms, however,
which are the leading order terms of neutral pion production, produce very
different results. Due to the violation of chiral symmetry, and therefore
PCAC, also the LET of pion photoproduction are violated. This violation is
proportional to the anomalous magnetic moments $\kappa$ in the Pauli current.
For a Dirac particle without anomalous magnetic moment the chiral symmetry
violation would not become visible in pion photoproduction.

As shown by Adler \cite{Adl70}, the PCAC relation (\ref{mod7}) will be
modified by higher order contributions due to anomalies. Most famous is the
triangle anomaly that explains the decay $\pi^0 \rightarrow 2 \gamma$, but
also square or pentagon etc. anomalies are possible with always an odd
number of axial vertices. Fig. 9g,h show such contributions for pion
production. Due to the weak e.m. coupling, the numerical value of the
anomaly with photon exchange is very small. However, larger contributions
arise if the exchanged photon is replaced by vector mesons, $\omega$ and
$\rho$. In practical calculations the fermion loops can be contracted to
effective couplings with coupling strengths extracted from radiative decays
of vector mesons.

The interaction Lagrangians for vector meson exchange are
\begin{eqnarray} \label{mod9}
\scr L_{\gamma\pi V} & = & \frac{g_{\gamma\pi V}}{m_{\pi}}\epsilon_{\mu\nu
\rho\sigma}\partial^{\mu}A^{\nu}{\bs \Phi}\cdot\partial^{\rho}
V^{\sigma}, \\
\scr L_{VNN} & = & -\bar\psi\left(g_{V1}\gamma_{\mu}-\frac{g_{V2}}{2m_{N}}
\sigma_{\mu\nu}\partial^{\nu}\right)V^{\mu}\psi.
\end{eqnarray}
The dominant contributions of vector meson exchange in pion production is
obtained in $E_{0+}$ and $M_{1-}$ multipoles for $\gamma, \pi^0$ with
$\omega$ exchange roughly a factor of 3 larger than $\rho$ exchange due to
$g_{\gamma\pi\rho}=g_{\gamma\pi\omega}/3$.

Besides the field theoretical models, pion photoproduction has also been
studied in quark models, particularly in the threshold region. In the
constituent quark model it is possible to derive LET by using PV coupling
between quarks and pions. In such a calculation, the spurious $cm$ motion
of the $qqq$ system can be avoided by taking harmonic oscillator wave
functions \cite{Dre84a}. However, this is not possible in relativistic bag
models, where spurious $cm$ motion can be suppressed only in approximate
ways \cite{Sch87,Kon91}. Also the Skyrme model does not reproduce LET due
to shortcoming of the model in the pion-nucleon coupling and due to the
neglect of the Dirac sea in the theory \cite{Sch91}.

Finally, due to the suppression of neutral pion production at threshold,
the rescattering graph of fig, 9i may become important. It involves the
production of a positive pion followed by charge exchange. This results in
a cusp effect at the $\pi^+$ threshold with a rapid variation of the s-wave
amplitude. Such an effect can be studied either in the K-matrix approach
\cite{Dal60} or
in dynamical models. Nozawa, Blankleider and Lee \cite{Noz90} have derived
a coupled-channels calculation of $\pi N$ scattering and photoproduction in
non-relativistic scattering theory. Fitting separable potentials to the
$\pi N$ phase shifts they obtained a reasonable description of pion
photo- and electroproduction up to about 500 MeV, see figs. 7 and 8. The
advantage of such coupled-channels calculations is that Watson's theorem
is exactly fulfilled in all partial waves.

\begin{center}
{\bf 5. Threshold pion photoproduction}
\end{center}

Pion photoproduction at threshold can be calculated in current algebra
using e.m. current conservation and PCAC hypothesis. Unlike Thomson
scattering, the best example for an exact low energy theorem, the LET for
pion photoproduction are evaluated as a power series in $\mu = m_\pi/m_N$.
In isospin decomposition we find \cite{Dre92}
\begin{eqnarray}
E_{0+}^{(-)} & = & \frac{e f_{\pi NN}}{4\pi m_\pi} \frac{1}{1+\mu}
 (1 + O(\mu^2) ) \nonumber\\
E_{0+}^{(+,0)} & = & \frac{e f_{\pi NN}}{4\pi m_\pi} \frac{1}{1+\mu}
 (-\frac{1}{2}\mu + \frac{1}{2}\mu^2 \mu_N^{(V,S)} + O(\mu^3) ),
\end{eqnarray}
where $\mu_N^{(S,V)}=\frac{1}{2} (\mu_p\pm\mu_n)=(0.44, 2.35)$
are the isoscalar and isovector magnetic moments of the nucleon. Expressed
in the four physical channels (28) we obtain
\begin{eqnarray}
E_{0+} (\gamma p \rightarrow n \pi^+) & = & \frac{e f_{\pi NN}}{4\pi m_\pi}
       \frac{1}{1+\mu} \sqrt{2} (1 - \frac{1}{2}\mu + O(\mu^2))\nonumber\\
E_{0+} (\gamma n \rightarrow p \pi^-) & = & \frac{e f_{\pi NN}}{4\pi m_\pi}
       \frac{1}{1+\mu} \sqrt{2} (-1 - \frac{1}{2}\mu + O(\mu^2))\nonumber\\
E_{0+} (\gamma p \rightarrow p \pi^0) & = & \frac{e f_{\pi NN}}{4\pi m_\pi}
       \frac{1}{1+\mu} (-\mu + \mu^2 \mu_p + O(\mu^3))\nonumber\\
E_{0+} (\gamma n \rightarrow n \pi^0) & = & \frac{e f_{\pi NN}}{4\pi m_\pi}
       \frac{1}{1+\mu} (-\mu^2 \mu_n + O(\mu^3))
\end{eqnarray}
\begin{table}[htbp]
\begin{center}
\renewcommand{\arraystretch}{1.25}    
\begin{tabular}{cccccc}\hline
Channel & LET & PV & PV+$\omega,\rho,\Delta$ & CQM & Experiment \\
\hline
$\gamma p \rightarrow n \pi^+$ & 27.5 & 27.7 & 27.7 & 29.0 & 28.6$\pm$0.2 \\
$\gamma n \rightarrow p \pi^-$ &-32.0 &-31.9 &-31.6 &-33.3 &-31.5$\pm$1.0 \\
$\gamma p \rightarrow p \pi^0$ & -2.4 & -2.5 & -2.3 & -2.7 & -2.2 \\
$\gamma n \rightarrow n \pi^0$ &  0.4 & 0.4  &  0.4 &  0.4 &       \\
\hline
\end{tabular}
\end{center}
{\small {\bf Table 3:}
Pion photoproduction amplitude $E_{0+}$ in units of
$10^{-3}/m_{\pi^+}$. LET: low energy theorems; PV: Born terms in
pseudovector coupling; PV+$\omega,\rho,\Delta$: pseudovector Born terms
plus t-channel vector meson exchange and u-channel $\Delta$-resonance;
CQM: constituent quark model \cite{Dre84a}. Experimental data for
$\gamma, \pi^\pm$ from Adamovitch \cite{Ada76}; for $\gamma, \pi^0$
on the proton threshold extrapolation from ref. \cite{Dre92}. }
\end{table}

In table 3 we show the predictions of the LET from eq.~(47) together with
the pseudovector Born terms and corrections from vector mesons and
$\Delta(1232)$. In addition we give the results of the constituent quark
model that fulfills the low energy theorems. A comparison with the
experiment gives an overall good agreement. For $\gamma p \rightarrow
p \pi^0$ we have listed the threshold extrapolation we obtained in an
analysis of the Mainz experiment by Beck et al \cite{Bec90}. Fig. 10a gives
the total cross section of the Mainz and the Saclay experiments and Fig. 10b
shows the angular distribution at 149.1 MeV, where the data differs
significantly from the LET result.

Expressed in s- and p-wave
multipoles and neglecting terms that are quadratic in the small p-wave
mles $E_{1+}$ and $M_{1-}$, the differential cross section is given by
\begin{eqnarray} \label{sigABC}
\frac{d\sigma}{d\Omega} & = & \frac{k}{q}(A + B cos \theta + C cos^2 \theta)\\
A & = & |E_{0^+}|^2 + \frac{5}{2} |M_{1^+}|^2
- Re\{M_{1+}^* (3 E_{1+}-M_{1^-})\}\nonumber\\
B & = & 2 Re\{E_{0+}^* (M_{1+}+3E_{1+}-M_{1^-})\}\nonumber\\
C & = & - \frac{3}{2} |M_{1^+}|^2 + 3Re\{M_{1+}^* (3 E_{1+}-M_{1^-})\}.
\nonumber
\end{eqnarray}
For the total cross section all interfences between different multipoles
disappear and we obtain the following closed form
\begin{eqnarray}
\sigma_{tot} & = & \int\frac{d\sigma}{d\Omega}d\Omega \nonumber\\
& = & 2\pi\frac{k}{q}\sum_{l=0}^{\infty}(l+1)^2[(l+2)(\mid E_{l+}\mid^2
+\mid M_{l+1,-}\mid^2)+l(\mid M_{l+}\mid^2+\mid E_{l+1,-}\mid^2)]\nonumber\\
& = & 4\pi\frac{k}{q}(\mid E_{0+}\mid^2+2\mid M_{1+}\mid^2\dots).
\end{eqnarray}

\begin{figure}[htbp]
\vspace{8cm}
\vspace{0.4cm}
{\small {\bf Fig. 10:}
{\bf (a)} Total and {\bf (b)} differential cross section at
$\omega=149.1 MeV$ for $p(\gamma, \pi^0)p$. The full lines show the result
of the fit of ref. \cite{Dre92} and the dashed lines are obtained for a
constant $E_{0+}=-2.4\cdot 10^{-3}/m_\pi$ (LET value) and the p-wave
multipoles of ref. \cite{Dre92}. The experimental points are from the Mainz
experiment ($\circ$) by Beck et al \cite{Bec90} and from the Saclay
experiment ($\Box$) by Mazzucato et al \cite{Maz86}.}
\end{figure}

In fig. 11 we show the threshold amplitudes for $p(\gamma, \pi^0)p$
determined from the cross section data of the Mainz experiment, Beck et al
\cite{Bec90}.The error band for the M1 amplitude and the five solid points are
from the analysis of Drechsel and Tiator \cite{Dre92}. The open circles show
the $E_{0+}$ multipole obtained from the total cross sections, when the
p-wave contributions \cite{Dre92} are subtracted.

\begin{figure}[htbp]
\centerline{\parbox{16cm}{\psfig{figure=ali11.ps,height=6.5cm}}}

\vspace{0.2cm}
{\small {\bf Fig. 11:}
Threshold s- and p-wave multipoles for $p(\gamma, \pi^0)p$. The p-wave
error band and the full dots show the combined analysis of differential
and total cross sections \cite{Dre92} obtained from the data of
Beck et al \cite{Bec90}. The open circles show our analysis obtained
from the total cross
sections by subtracting the p-wave amplitudes of ref \cite{Dre92}. The error
bars give the total correlated errors from the cross sections and
from the determination of the p-wave multipoles.
The dotted and full lines show the results of chiral perturbation theory
\cite{Ber92,Ber92a} for isospin symmetry and isospin breaking due to the
physical pion and nucleon masses respectively.
}
\end{figure}

Close to the $\gamma, \pi^0$ threshold the data approaches the prediction
of the LET, around the $\gamma, \pi^+$ threshold, however,
the amplitude is consistent with zero. In fig. 10b we also show the
calculations of Bernard et al \cite{Ber92,Ber92a,Mei93} in chiral
perturbation theory.
While a consistent 1-loop calculation \cite{Ber92} assuming isospin symmetry
gives a threshold value of $-1.33\cdot 10^{-3}/m_{\pi^{+}}$,
an approximate treatment of isospin
breaking due to different pion and nucleon masses also calculated in
1-loop gives a cusp effect similar to K-matrix calculations and a shape
closer to the data \cite{Ber92a}.

While the differential cross section is very sensitive to the asymmetry
parameter $B$ in eq. (\ref{sigABC}), and therefore to the real parts of
$E_{0+}$ and
the p-wave combination M1, a complete analysis of the threshold region
with separation of the individual p-wave multipoles and the imaginary part
of the s-wave can only be obtained with polarization observables.
In the same approximation as in eq. (\ref{sigABC}) the three single
polarization observables are given by
\begin{eqnarray}
\Sigma & = & \frac{3 k sin^2 \theta}{2 q d\sigma/d\Omega}
(|M_{1+}|^2 + 2 Re\{M_{1+}^* (E_{1+}+M_{1-})\}\\ \nonumber
T & = & \frac{3 k sin \theta}{q d\sigma/d\Omega}
Im\{E_{0+}^* (E_{1+}-M_{1+}) - cos\theta M_{1+}^* (4 E_{1+}+M_{1-})\}\\
\nonumber
P & = & -\frac{k sin \theta}{q d\sigma/d\Omega}
Im\{E_{0+}^* (3E_{1+}+M_{1+}+2M_{1-}) + 3 cos\theta M_{1+}^* M_{1-}\}.
\end{eqnarray}
Complete formulas without approximations from the cross section and the
polarization observables can be found in the appendix of ref. \cite{Dre92}.

\begin{figure}[htbp]
\centerline{\parbox{16cm}{\psfig{figure=ali12.ps,height=6.5cm}}}

\vspace{0.2cm}
{\small {\bf Fig. 12:}
Target asymmetry (T) and photon asymmetry ($\Sigma$) for threshold
$p(\gamma, \pi^0)p$ at $\theta_{c.m.}=90^\circ$. In {\bf a)} the
solid line shows
the full calculation and the dash-dotted line the situation for
$Im E_{0+}$=0. The dotted curve is a magnification of the full line by
a factor of 100 in the region between the neutral and the charged pion
thresholds. In {\bf b)} the full, dash-dotted and dotted lines are obtained
for ($E_{1+}$, $M_{1-}$) = (-0.5, -2.3), (0, -0.8) and (0.3, 0)
respectively. Note that all three sets give 3$E_{1+}$-$M_{1-}$=0.8 in
agreement with the analysis shown in fig. 11. (All p-wave multipoles are
given in units of $q k \cdot 10^{-3}/m_{\pi^+}^3$)
}
\end{figure}
The target and recoil polarizations are very sensitive to the imaginary
part of the s-wave amplitude. The beam asymmetry gives together with
the differential cross section another linear independent measurement of
the small p-wave multipoles. In fig. 12 we demonstrate these effects with
a calculation done by Davidson \cite{Dav91}. The target asymmetry is
directly
proportional to $Im E_{0+}$ and shows a very steep rise at the charged
pion threshold. As also pointed out by Bernstein \cite{Ber93}, a measurement of
this quantity provides the only possibility to measure directly the $\pi N$
scattering lengths which is a very sensitive test to chiral dynamics.
In the region below the charged pion threshold there
is even a possibility to observe the isoscalar scattering length for
$\pi^0 p$ scattering. However due to the smallness of this amplitude,
the target asymmetry will be around a hundred times smaller than in the
region where the isovector amplitude dominates. From the small isoscalar
scattering length not even the sign is known, the calculation presented
here assumes a negative sign and a value of $a(\pi^0 p)=-.005/m_\pi$.
In fig. 12b we show the possibility to disentangle the small p-wave
multipoles, all three curves would be totally consistent with the
differential cross section data.

\begin{center}
{\bf 6. Pion photoproduction at resonance}
\end{center}

As can be seen in figs. 1 and 2, the $\Delta$-resonance plays a dominant
role in pion photoproduction, for neutral pions even very close to
threshold. Fig. 9e,f show the s- and u-channel Feynman diagrams in analogy
to the Born terms with nucleon poles. However, the off-shell extrapolation
of propagators and vertices gives rise to large model dependences for spin
3/2 particles. Therefore in most calculations the resonances are calculated
on-shell with only s-channel contributions. Furthermore the propagators are
split into forward and backward propagating pieces neglecting the backward
propagating antiparticle contribution. In table 4 we list the most
important nucleon resonances that can be excited in photoproduction of
pions and etas. In pion production, the $\Delta(1232), N^*(1520)$ and
$N^*(1680)$ are the dominant candidates of the first,
second and third resonance
region, see fig. 1, and in eta production the $N^*(1535)$ plays a similarly
dominant role as the $\Delta$ for the pion. Due to isospin, delta
resonances ($I=3/2$) cannot decay in etas ($I=0$), but also in pion production
contributions from higher deltas are small.
\begin{table}[htbp]
\begin{center}
\renewcommand{\arraystretch}{1.5}    
\begin{tabular}{cc|ccc|cc}\hline
Resonances & $I(J^{P})$ & $N\pi$ & $ N\eta$ & $N\pi\pi$ & $p\gamma$ & $n\gamma$
\\
\hline
$P_{33}(1232)$ & $\frac{3}{2}\,(\frac{3}{2}^{+})$  & 99.4 &  &  & 0.56-0.66 &
0.56-0.66 \\
$P_{11}(1440)$ & $\frac{1}{2}\,(\frac{1}{2}^{+})$  & 60-70 &  & 30-40 &
0.08-0.10
& 0.01-0.06  \\
$D_{13}(1520)$ & $\frac{1}{2}\,(\frac{3}{2}^{-})$  & 50-60 & $\sim$ 0.1 & 40-50
&
0.43-0.57 & 0.34-0.51 \\
$S_{11}(1535)$ & $\frac{1}{2}\,(\frac{1}{2}^{-})$  & 35-55 & 30-50 & 5-20 &
0.1-0.2 & 0.15-0.35\\
$S_{11}(1650)$ & $\frac{1}{2}\,(\frac{1}{2}^{-})$  & 60-80 & $\sim$ 1& 5-20 &
0.04-0.16 & 0-0.17 \\
$D_{15}(1675)$ & $\frac{1}{2}\,(\frac{5}{2}^{-})$  & 40-50 & $\sim$ 1& 50-60 &
$\sim$ 0.01 & 0.07-0.12 \\
$F_{15}(1680)$ & $\frac{1}{2}\,(\frac{5}{2}^{+})$  & 60-70 &   & 30-40&
0.21-0.30 & 0.02-0.05 \\
\hline
\end{tabular}
\end{center}
{\small {\bf Table 4:}
Branching rations of the most important nucleon resonances for
$\pi$ and $\eta$ photoproduction \cite{PDG92}.}
\end{table}

For electroproduction the situation is not clear and there is very
little information available on longitudinal excitations of nucleon resonances.
Possible candidates are $C2$ of $\Delta(1232)$, $C0$ of $N^*(1440)$ and
$C1$ of $N^*(1535)$. For further reading, see refs. \cite{Dre93,Bur91}.

\begin{figure}[htbp]
\vspace{10cm}
{\small {\bf Fig. 13:}
Differential cross section and photon asymmetry ($\Sigma$) for the
reactions $p(\gamma,\pi^0)p$ (a,b) and $p(\gamma,\pi^+)n$ (c,d)
in the resonance region.
The calculation are done by Davidson et al \cite{Dav91} in a unitarized
field theoretical model. In (c,d) the dashed line gives the contributions of
s- and p-wave multipoles only, which is close to the full calculations in
$(\gamma,\pi^0)$. The data are from ref. \cite{Men77}. }
\end{figure}

Fig. 13 shows a calculation by Davidson et al \cite{Dav91} for pion
photoproduction in the resonance region. The model is fully unitarized and
takes into account Born terms, vector mesons and the $\Delta$-resonance
$M1$ and $E2$ excitation. A very good agreement is obtained with
experimental data of \cite{Men77} for both differential cross section
and photon asymmetry. Currently a new experiment is under way by Beck et al
\cite{Bec92} at Mainz proposing to measure those four observables with
high enough accuracy in order to get a precise determination of the $E2$ to
$M1$ ratio of the $\Delta$ excitation. The two channels $(\gamma,\pi^0)$ and
$(\gamma,\pi^+)$ on the proton are both necessary to get a separation of the
isospin 3/2 contribution of the p-wave multipoles. The analysis is
simplified by the fact that the cross section and the beam asymmetry for
$(\gamma,\pi^0)$ are mainly given by s- and p-wave multipoles with only small
contributions from higher multipoles. This is more difficult in the case
of $(\gamma,\pi^+)$, where the higher multipoles make all the difference
between the dashed and full curves in figs. 13c,d. However, this is only
due to the pion pole term (fig. 9c), and the precise knowledge of this
Born term contribution will be very helpful in the analysis.

A precise determination of $E2/M1$ (EMR ratio) which, expressed in
multipoles is equal to $E_{1+}/M_{1+}$, is a very stringent test of
isobar models. In the constituent quark model, it is predicted
from the color hyperfine interaction to be -1.5\% and can be directly related
to
the quadrupole deformation of the nucleon and the $\Delta$. Other quark model
calculations predict $-2\% \le EMR \le -1\%$, while the Skyrme model
obtains $-7\% \le EMR \le -4\%$. Analyses of the present experimental data
are in the range of $-5\% \le EMR \le -1.5\%$ with the smallest error bar
given in the analysis of Davidson et al \cite{Dav91} with $-1.57 \pm 0.72\%$,
see also fig. 13. Further details and references can be found in ref.
\cite{Dre93}.

\begin{center}
{\bf 7. Eta photoproduction}
\end{center}

With the recent completion of the modern
electron accelerators at Mainz and Bonn and the construction of new
detectors it is now possible to measure eta photoproduction from threshold
at 707 MeV up to 850 MeV at Mainz and even higher energies at Bonn with
a similar precision as pion photoproduction. A large amount of data has
already been taken and is currently being analysed\cite{Ant91,Kru93}.
These upcoming results will improve our knowledge of the $(\gamma,\eta)$
process enormously; currently it is only based on some very old
measurements of 20
years ago\cite{Del67,Bac68} and some more recent data from Tokyo\cite{Hom88}
and Bates\cite{Dyt90}.

Unlike pion photoproduction, no $LET$ can be derived for eta photoproduction
for 3 good reasons: ({\it i}) The expansion parameter
$\mu=m_{\eta}/m_{N}\approx
0.6$ is too large to provide convergence up to order $\mu^2$; ({\it ii})
due to large $\eta-\eta'$ mixing with a mixing angle of about $20^\circ$ and a
non-conserved axial singlet current $A_0^{\mu}$ for the $\eta'$, there is no
$PCAC$ theorem for eta mesons; ({\it iii}) there are nucleon resonances,
mainly the $S_{11}(1535)$ close at threshold ($W_{thr.}=1486$ MeV)
strongly violating the condition that the internal excitation energy must be
larger than the mass of the meson.

Therefore, it is not surprising that nucleon resonance excitation is the
dominant reaction process in $(\gamma,\eta)$. Firstly, in contrast to pions
which will excite $\Delta(T=3/2)$ as well as $N^*(T=1/2)$ resonances, the
$\eta$ meson will only appear in the decay of  $N^*$ resonances with $T=1/2$.
In the low-energy region this is dominantly the $S_{11}(1535)$ state that
decays 45-55\% to $\eta N$, the only nucleon resonance with such a
strong branching ratio to the $\eta$ channel. This result is even more
surprising as a near--by resonance of similar structure,
the $S_{11}(1650)$ has a
branching ratio of only 1.5\%. This `$\eta$ puzzle' is not yet understood
in quark models of the nucleon.

Our approach is a field theoretical model, taking into account nucleon Born
terms and vector meson exchange. The nucleon resonances $P_{11}(1440),\,
D_{13}(1520)$ and $S_{11}(1535)$ which play an important role in eta
photoproduction (see table 4) are described in the dynamical model of
Bennhold and Tanabe \cite{Ben91}.
Assuming an isobar model for each partial wave the transition amplitude can
be written as
\begin{eqnarray} \label{eta1}
t_{ij}(W)=f_{i}^{\dagger}D^{-1}(W)f_{j},
\end{eqnarray}
where $W$ is the invariant energy and $i,j=\pi,\eta$ denotes the $\pi N$ and
$\eta N$ channels, respectively. The vertex functions $f_i$ are parametrized
with coupling strengths and formfactors and the $N^*$ propagators are given
by
\begin{eqnarray} \label{eta2}
D(W)=W-m_0-\Sigma_{\pi}(W)-\Sigma_{\eta}(W)+\frac{i}{2}\Gamma_{\pi\pi}(W)
\end{eqnarray}
with the bare resonance mass $m_0$.

The self-energy $\Sigma$ associated with the $\pi N$ and $\eta N$
intermediate states is given by
\begin{eqnarray} \label{eta3}
\Sigma_{i}(W)=\int_0^{\infty}\frac{q^2dq}{(2\pi)^3}\,\frac{M}{2w_i(q)E_N(q)}
\left(\frac{q}{m_i}\right)^{2l}\frac{g_i^2(1+q^2/\Lambda_i^2)^{-2-l}}
{W-w_i(q)-E_N(q)+i\epsilon}
\end{eqnarray}
with $w_i(q)=\sqrt{m_i^2+q^2},\,E_N(q)=\sqrt{M_{}^2+q^2}$ and $M$ denoting
the nucleon mass. The $2\pi$--decay width $\Gamma_{\pi\pi}$ is parametrized
with one free parameter. The six parameters in this approach have been
determined  for each partial wave by a least-squares fit to all data of the
reactions  $\pi N \rightarrow \pi N,\,\pi N \rightarrow \pi\pi N$ and
$\pi^- p \rightarrow \eta n$ and can be found in ref.\cite{Ben91}

For a convenient use of this operator, especially in nuclear applications
with multidimensional integrals, we have obtained simple parametrizations of
the self-energy $\Sigma$, eq.~(\ref{eta3}), in very good agreement with the
exact
numerical values.
\begin{eqnarray} \label{eta4}
Re\,\Sigma&=&a + (b_1\sqrt{x}+b_2 x^2)\Theta(-x) + (c_1 x+c_2 x^2)\Theta(x)\,,
\nonumber\\
I\!m\,\Sigma&=&(d_1\sqrt{x}+d_2 x+d_3 x^2)\Theta(x)
\end{eqnarray}
with $x=(W-M-m_i)/m_{\pi}$, $i=\pi,\eta$ and the step function
$\Theta(x)$. The parameters are given in table 5.
\begin{table}[htbp]
\begin{center}
\renewcommand{\arraystretch}{1.5}    
\begin{tabular}{cc|cccccccc}\hline
& & $a$ & $b_1$ & $b_2$ & $c_1$ & $c_2$ & $d_1$ & $d_2$ & $d_3$ \\ \hline
$S_{11}$ & $\pi N$ & 17 & 0 & 0 & 0 & 0 & -129.5 & 80 & -5 \\
        & $\eta N$ & -27 & 17.7 & -1.23 & 22.9 & -5.17 & -38.1 & 18.3 & 0
\\[0.2cm]
$P_{11}$ & $\pi N$ & -150 & 0 & 0 & 0 & 0 & 55.1 & -96.2 & 6.6 \\[0.2cm]
$D_{13}$ & $\pi N$ & -26 & 0 & 0 & 0 & 0 & 23 & -32.1 & 2.7 \\ \hline
\end{tabular}
\end{center}
\begin{center}
{\small {\bf Table 5:}
Parameters for the $\pi N$ and $\eta N$ self-energies in $MeV$}
\end{center}
\end{table}
Finally the decay width in the 2$\pi$--channel is given by
\begin{eqnarray} \label{eta5}
\Gamma_{\pi\pi}(W)=\gamma\, x\, \Theta(x)\,,\qquad x=(W-M-2m_{\pi})/m_{\pi}
\end{eqnarray}
with $\gamma(S_{11})=4.3 MeV$, $\gamma(P_{11})=80.3 MeV$ and
$\gamma(D_{13})=24.2 MeV$.

With the hadronic vertex and propagators being determined, the
photoproduction amplitudes for $(\gamma,\pi)$ and $(\gamma,\eta)$
are given by
\begin{eqnarray} \label{eta6}
t_{i\gamma}(W)\,=\,V_{i\gamma}^B(W)\,+\,f_{i}^{\dagger}D^{-1}(W)
\tilde{f_{\gamma}}\,,
\end{eqnarray}
where $V_{i\gamma}^B$ are the Born terms and $\tilde{f_{\gamma}}$ the
electromagnetic vertex. The latter was determined by using the pion
photoproduction data. In this way there are no free parameters left
for the $(\gamma,\eta)$ process which will turn out as a prediction
rather than a
fit in this model. Since in ref.\cite{Ben91} the Born terms  in the
$\eta$--channel have been neglected, $V_{\eta\gamma}^B\equiv0$, this model
consists out of four $(\gamma,\eta)$ multipoles only: From $S_{11}(1535)$ the
strongly dominating $E_{0+}$, from $P_{11}(1440)$ the $M_{1-}$ and from
$D_{13}(1520)$ the $E_{2-}$ and $M_{2-}$.

Whereas the neglect of the $(\gamma,\eta)$ Born terms are within the
uncertainties of the older experimental data for the proton,
they can play a more important role when better data will become available and
will become absolutely necessary in nuclear reactions like the coherent $\eta$
photoproduction on $^4$He, where the dominant excitation of the $S_{11}$
resonance is forbidden. The evaluation of the background terms is
straightforward and in complete analogy to $(\gamma,\pi^0)$ except to the fact
that the $\eta$ is an isoscalar meson. Furthermore, the pseudoscalar
coupling of the $\eta N$ is not ruled out by $LET$ as in the case of
$(\gamma,\pi)$, in fact, as we will show later, present experimental data are
in favour of the $PS$ rather than the $PV$ coupling.

The effective Lagrangians for $\eta N$ coupling are given by
\begin{eqnarray}  \label{eta7}
{\scr L}_{\eta NN}^{PS}=-ig_{\eta NN}\bar{\psi}\gamma_5\psi\phi_{\eta}\,,
\qquad
{\scr L}_{\eta NN}^{PV}=\frac{g_{\eta NN}}{2M}\bar{\psi}\gamma_{\mu}
\gamma_5\psi\partial^{\mu}\phi_{\eta}\,.
\end{eqnarray}
With the electromagnetic Lagrangian
\begin{eqnarray} \label{eta8}
{\scr L}_{\gamma NN}^{PS}=-e\bar{\psi}\gamma_{\mu}\frac{1+\tau_0}{2}
\psi A^{\mu}+\frac{e}{4M}\bar{\psi}(\kappa^S+\kappa^V\tau_0)\sigma_{\mu\nu}
\psi\,F^{\mu\nu}\,,
\end{eqnarray}
where $\kappa^S=-0.06$ and $\kappa^V=1.85$ are the isoscalar and isovector
anomalous magnetic moments and
$F^{\mu\nu}=\partial^{\nu}A^{\mu}-\partial^{\mu}A^{\nu}$
we can evaluate the $s$-- and $u$--channel Born terms.

Due to the decay of the vector mesons ${\mrm V}(J^{\pi};T)=\omega(1^-;0)$ and
$\rho(1^-;1)$ into $\eta\,\gamma$ we also have to include the $t$--channel
Born diagrams which we evaluate from the Lagrangians
\begin{eqnarray} \label{eta9}
{\scr L}_{{\mrm V}NN}=-g_{V_1}\bar{\psi}\gamma_{\mu}\psi V^{\mu}+
\frac{g_{V_2}}{4M}\bar{\psi}\sigma_{\mu\nu}\psi\,V^{\mu\nu}\,,
\qquad
{\scr L}_{{\mrm V}\eta\gamma}=\frac{e\lambda_{\mrm V}}{4m_{\eta}}
\varepsilon_{\mu\nu\lambda\sigma}F^{\mu\nu}V^{\lambda\sigma}\phi_{\eta}
\end{eqnarray}
with
$V^{\mu\nu}=\partial^{\nu}V^{\mu}-\partial^{\mu}V^{\nu}$
like the e.m. field tensor $F^{\mu\nu}$.

The isospin decomposition of the eta photoproduction amplitudes gives only
2 amplitudes because of isospin $I=0$ of the $\eta$ meson. So we can
define the isoscalar and isovector
amplitudes $F_i^{(0)}$ and $F_i^{(1)}$ by
\begin{eqnarray} \label{eta10}
F_i=F_i^{(0)}+F_i^{(1)}\tau_0
\end{eqnarray}
or similarly the multipoles $E_{l\pm}$ and $M_{l\pm}$.
Due to the isospin, the $\omega$ contributes only to $F_i^{(0)}$ and $\rho$
only to $F_i^{(1)}$.

\begin{table}[htbp]
\begin{center}
\renewcommand{\arraystretch}{1.5}    
\begin{tabular}{c|cccc}\hline
${V}$ & $g_{V_1}^2/4\pi$ & $g_{V_2}/g_{V_1}$ & $\Lambda_{V} (MeV)$ &
$\lambda_{V}$
\\ \hline
$\omega$   &  23  &  0  & 1400 & 0.192 \\
$\rho$     &  0.5 & 6.1 & 1800 & 0.89 \\ \hline
\end{tabular}
\end{center}
{\small {\bf Table 6:}
Coupling constants and cut--off masses for
the background vector meson exchange contributions.}
\end{table}

In table 6, we give the coupling constants and cut--off masses for the
background contributions. For the vector mesons we have introduced dipole
formfactors
$F(\bs{k}^2)=(\Lambda_{V}^2-m_{V}^2)^2/(\Lambda_{V}^2+
\bs{k}^2)^2$ on the
${\mrm V} NN$ vertex given by the Bonn potential, for the $\eta NN$ coupling
the formfactors turned out to be insensitive and have been ignored. The e.m.
V$\eta\gamma$ couplings are obtained from the partial decay widths of the
vector mesons. The largest uncertainty, however, appear in the $\eta NN$
coupling. Here not even the structure of the coupling $PS$ or $PV$ is known.
For the coupling constant $g_{\eta NN}^2/4\pi$ the values in the literature
range
between 1 and 7, the larger ones are found in the Bonn potential\cite{Bro90}
while the smaller values are preferred by the current $(\gamma,\eta)$ data on
the proton. An important aim of the eta photoproduction will
also be a better determination on this coupling constant which is rather
insensitive in $NN$ interaction.\\

\begin{table}[htbp]
\begin{center}
\renewcommand{\arraystretch}{1.5}    
\begin{tabular}{c|ccccc}\hline
target & $S_{11}(1535)$ & $\omega$ & $\rho$ & Born $PS(1.0)$ & Born $PV(1.0)$
\\
\hline
proton   & 12.91 + 5.97 i & 0.35 &  2.63 & -5.20 & -0.88 \\
neutron  & -7.12 - 4.86 i & 0.35 & -2.63 &  3.55 & -1.04 \\
\hline
\end{tabular}
\end{center}
{\small {\bf Table 7:}
Contributions to the threshold amplitudes
of $E_{0+}$ in units of $10^{-3}/m_{\pi}$. In the lab frame, the threshold
photon energies are $707.16 MeV$ on protons and $706.94 MeV$ on neutrons.}
\end{table}

In table 7 we give the individual contributions to the threshold $E_{0+}$
multipole for protons and neutrons. While the $S_{11}$ contribution is
complex, the background contribution from Born terms and vector mesons is
real. In the table, the Born PS and PV terms are calculated for a coupling
of $g_{\eta NN}^2/4\pi = 1$ and scales proportional to $g_{\eta NN}$. Note
that for a coupling of $g_{\eta NN}^2/4\pi = 0.33$ the background vanishes
in a pseudoscalar model. Such a coupling strength seems to be favoured by
preliminary results of the Mainz experiment \cite{Kru93}.

In contrast to the $\pi N$ interaction, little is known about the $\eta N$
interaction and, consequently,
about the $\eta NN$ vertex.  In the
case of pion scattering and pion photoproduction the $\pi NN$ coupling
is preferred to be pseudovector (PV), in accord with current algebra
results and chiral symmetry.  However, because the eta mass is so much
larger than the pion mass leading to large SU(3)
breaking, and even more due to the $\eta-\eta'$ mixing with the
non-conserved axial singlet current $J_{5\mu}^0$, there is no
compelling reason to select the PV rather than
the PS form for the $\eta NN$ vertex.  We therefore explore both
possibilities below in the hope that the new experimental data from Mainz
and Bonn will be able to distinguish between the different forms.

The uncertainty in the structure of the $\eta NN$ vertex is equally
large regarding the magnitude of the coupling constant. In SU(3) the
coupling constants of $\pi, K$ and $\eta$ are related by
\begin{equation} \label{sucoup}
\frac{g_{\eta_8}^2}{4 \pi} = \frac{1}{3}(3-4\alpha)^2 \frac{g_{\pi}^2}{4 \pi}.
\end{equation}
Depending on the data avilable in $YN$ scattering,
the $\eta$ coupling ranges between
$0.5-1.7$.
Other determinations of the $\eta N$
coupling employ reactions involving the eta, such as
$\pi^- p \rightarrow \eta n$, and range from 0.6 - 1.7.  Finally, the
$\eta$-meson has also been built into one boson exchange potentials
(OBEP) of the nucleon-nucleon interaction.  Typical values obtained in
fits with the Bonn potential can lie anywhere between 3 - 7 \cite{Bro90}.
However,
including the $\eta$ yields only small effects in fitting the $NN$ phase
shifts and, furthermore, provide a small contribution to nuclear
binding at normal nuclear densities.

\begin{figure}[htbp]
\centerline{\psfig{figure=ali14.ps,height=7.5cm}}

\vspace{0.2cm}
{\small {\bf Fig. 14:}
Total cross section for eta photoproduction on the
proton. The full lines are the result of the
$S_{11}$ resonance and vector meson contributions without Born terms and
the dashed lines show the effect of the additional Born terms with
different coupling constants for both PS ($g_{\eta NN}^2/4\pi=0.1, 0.5,
1.0, 3.0$) and PV ($g_{\eta NN}^2/4\pi=1.0, 3.0, 6.0, 10.0$).}
\end{figure}

In fig. 14 we show the total cross section for eta photoproduction on the
proton from threshold up to 780 MeV. The full lines are the result of the
$S_{11}$ resonance and vector meson contributions without Born terms and
the dashed lines show the effect of the additional Born terms with
different coupling constants for both PS and PV coupling. Since the total
cross section is practically only s-wave, the individual contributions are
as in table 7. The $E_{0+}$ amplitude is a factor of 6 smaller in PV
coupling compared to PS coupling, and therefore even very large coupling
constants as $g_{\eta NN}^2/4\pi = 10$ do not lower the cross section by
much. In a PS model, however, the range of couplings $g_{\eta NN}^2/4\pi
\le 3$ gives a very broad band of cross sections with the possibility
to determine the coupling strength very accurately.

\begin{figure}[htbp]
\centerline{\psfig{figure=ali15.ps,height=7cm}}

\vspace{0.2cm}
{\small {\bf Fig. 15:}
Total cross section for eta photoproduction on the
proton. The dotted and dash-dotted lines are the results of the
$N^*$ resonances alone and with vector meson contributions respectively.
The short dashed and long dashed curves are obtained with additional
PV Born terms ($g_{\eta NN}^2/4\pi=6.0$) or PS Born terms
($g_{\eta NN}^2/4\pi=0.1$). }
\end{figure}
\begin{figure}[htbp]
\centerline{\psfig{figure=ali16.ps,height=9.5cm}}

\vspace{0.2cm}
{\small {\bf Fig. 16:}
Differential cross section for eta photoproduction on the proton at
737, 752, 767 and 782 MeV photon lab energy. The notation of the curves
are as in fig. 15.}
\end{figure}
\begin{figure}[htbp]
\centerline{\psfig{figure=ali17.ps,height=9cm}}

\vspace{0.2cm}
{\small {\bf Fig. 17:}
Influence of the $P_{11}(1440)$ and $D_{13}(1520)$ resonances in the
differential cross section $d\sigma/d\Omega$ and in the single polarization
observables $\Sigma$, $T$ and $P$ at a photon lab energy of 752 MeV.
The full lines show the complete calculation with resonances, vector mesons
and PS Born terms with $g_{\eta NN}^2/4\pi=0.4$. The dashed and dash-dotted
lines are obtained when the $P_{11}$ or the $D_{13}$ resonances are
omitted.}
\end{figure}

However, with the total cross section data alone the analysis is not unique.
It is possible to get a similar total cross section, e.g. for a PS coupling
of 0.1 and a PV coupling of 6.0. This is demonstrated in fig. 15. Data that
fall below this curve can practically only be explained with a pseudoscalar
model. For example, a coupling strength of $g_{\eta NN}^2/4\pi=0.4$ would
give results very close to the pure resonance contribution (dotted line).
Large couplings around 1.0 or 1.4 discussed in eta photoproduction before
\cite{Muk91} would be consistent only with data considerably below 15
$\mu b$ at maximum. Much clearer, however, is the answer to the question
on PS vs. PV coupling by studying differential cross sections. In fig. 16
we compare the 2 models PS(0.1) and PV(6.0) which both give the same total
cross section. Clearly, the forward backward asymmetry is so different,
that the new Mainz experiment can easily answer this question. The
preliminary results of Krusche et al give a rather flat cross section
comparible with the PS model.

In future experiments polarization degrees of freedom will be possible and
will allow to search for effects of eta photoproduction that are hidden
in the dominant $S_{11}$ amplitude in the unpolarized cross section.
In fig. 17 we give the single polarization variables $\Sigma, T$ and $P$
together with the differential cross section at a photon energy of 752 MeV.
While the unpolarized cross section is rather insensitive to
resonances other than $S_{11}$, the photon asymmetry $\Sigma$ shows a large
sensitivity for the $D_{13}$ resonance, a somewhat smaller effect is also
present in the target polarization. The recoil polarization, finally, is
even sensitive to the $P_{11}$ Roper resonance that is very hard to detect
in electromagnetic reactions.

In a very recent experiment at Bonn, electroproduction of etas on the
proton has been measured \cite{Wil93}.
In this experiment the momentum transfer of the
virtual photons was rather small, $Q^2=0.056 GeV^2=1.4 fm^{-2}$, close to
the real photon limit. In fig. 18 we show the preliminary results together with
a
Breit-Wigner fit by Wilhelm \cite{Wil93}. In this fit a parametrization was
used with an energy dependent width $\Gamma(W)$ in the form
\begin{eqnarray} \label{et11}
\sigma_{tot} & = & \frac{k_\eta}{k_\gamma^{cm}}
\frac{A\,M_R^2\,\Gamma_R^2}{(M_R^2-W^2)^2 + M_R^2 \Gamma^2(W))}
\end{eqnarray}
with
\begin{eqnarray}
\Gamma(W) & = & \Gamma_R \left(0.50\frac{k_\eta}{k_\eta^R}
+ 0.40\frac{k_\pi}{k_\pi^R} + 0.10\right).\nonumber
\end{eqnarray}

\begin{figure}[htbp]
\centerline{\psfig{figure=ali18.ps,height=6.5cm}}

\vspace{0.2cm}
{\small {\bf Fig. 18:}
{\bf (a)} Total cross section for $e,e'\eta$ on the proton at $Q^2=0.056
GeV^2$.
The full line is the
result of a Breit-Wigner fit by Wilhelm \cite{Wil93}. {\bf (b)} Threshold
amplitude $\mid E_{0+}\mid$ under the asumption that longitudinal
amplitudes and higher partial waves are negligible. The full line is the
fit (\ref{fita}) as in a), the dashed line a fit with constant width
(\ref{fitb}).
The preliminary data are from a recent experiment at Bonn \cite{Wil93}. }
\end{figure}

The $cm$ momenta of $\pi$ and $\eta$ at energy $W$ are $k_\pi$ and $k_\eta$,
respectively. The momenta $k_\pi^R$ and $k_\eta^R$ are the $cm$ momenta at
resonance $W=M_R$, where the branching ratios $\Gamma_\eta/\Gamma_R=0.50$,
$\Gamma_\pi/\Gamma_R=0.40$ and $\Gamma_{\pi\pi}/\Gamma_R=0.10$ are used for
the $\eta N$, $\pi N$ and $\pi\pi N$ decay channels \cite{PDG92}. The
result of the fit in fig. 18 is \cite{Wil93}
\begin{eqnarray} \label{fita}
A  =  (41.57 \pm 0.75) \mu b, \quad
M_R  =  (1540.1 \pm 1.1) MeV, \quad
\Gamma_R =  (134.6 \pm 5.6) MeV,
\end{eqnarray}
in good agreement with the values of the Particle Data Group \cite{PDG92}:
\begin{eqnarray}
M_R  =  (1535 \pm 15) MeV, \quad
\Gamma_R  =  (150\,^{+100}_{-50}) MeV
\end{eqnarray}
for the $S_{11}(1535)$ resonance. It is interesting to note that the
occurance of a peak in the total cross section almost exactly at 1535 MeV
has nothing to do with a resonance peak. In fact, if we divide out the phase
space factor, the data in fig. 18b look very different and are
fall monotonically. This can be explained with the $\eta$ threshold very
close to the resonance pole, giving rise to a cusp effect and a strong
energy dependent width. In a simple Breit-Wigner analysis using constant
width, however, we obtain a resonance pole with
\begin{eqnarray} \label{fitb}
M_R  =  (1503 \pm 5) MeV, \quad
\Gamma_R  =  (142 \pm 14) MeV,
\end{eqnarray}
shown as the dashed line in fig. 18b.

\begin{figure}[htbp]
\vspace{9cm}
{\small {\bf Fig. 19:}
Breit-Wigner fit by Clajus and Nefkens \cite{Cla92} of selected data for
$\sigma_{tot}$ in the reaction $\pi^- p \rightarrow \eta n$.}
\end{figure}

Recently H\"ohler and Schulte \cite{Hoe93} pointed out that in elastic
pion scattering
the situation of the $S_{11}(1535)$ is rather obscure due to the close
$\eta$ threshold. In principle the peak around 1535 MeV could even be
described with a cusp at the $\eta$ threshold and a constant nonresonant
background. Only the $S_{11}(1650)$ showed a clear resonant behaviour.
In an analysis of Clajus and Nefkens \cite{Cla92} on $\pi^- p \rightarrow
\eta n$ total cross section data, fig. 19, they obtain a similar shape as
in electroproduction, however, quite different resonance parameters of
\begin{eqnarray} \label{fitc}
M_R  =  (1483 \pm 16) MeV, \quad
\Gamma_R  =  (204 \pm 21) MeV.
\end{eqnarray}
This fit was obtained with a constant width and has to be compared with
our fit (\ref{fitb}). However, it is based on a much more uncertain data base
compared to the electroproduction data.

As H\"ohler pointed out \cite{Hoe93} the peak structure of the total cross
section (figs.~18-19) is no signature for a resonance. At threshold the
cross section rises proportional to $k_\eta$ according to phase space and
at higher energies it falls proportional to $1/k_\eta^2$ by dimensional
argument and unitarity constraints. In order to get a clear answer a much
more precise determination of the threshold amplitudes will be necessary.
For the absolute value $\mid E_{0+}\mid$, this will be available after the
final analysis of the Mainz experiment. Using polarization degrees of
freedom, in future, it will even be possible to obtain a full
model independent analysis of both, the real and imaginary parts of the
photoproduction amplitudes.

\begin{center}
{\bf 8. Pion and eta photoproduction on light nuclei}
\end{center}

For light nuclei, pion and eta photoproduction can be calculated in a
coupled-channels framework. For the reaction of $^3He(\gamma,\pi^+)^3H$
this can be found in detail in refs. \cite{Kam91,Kam92}.
In momentum space the nuclear photoproduction amplitude can be written as
\begin{eqnarray} \label{cc1}
F_{\pi \gamma}(\bs{k},\bs{q}) = V_{\pi \gamma}(\bs{k},\bs{q}) -
\frac{a}{(2\pi)^2} \sum_{{\pi'}} \int \frac{d^3k'}{M(k')} \frac{F_{\pi
\pi'}(\bs{k},\bs{k}\,')V_{\pi'\gamma}(\bs{k}\,', \bs{k})}
{E(k) - E(k') + i \epsilon} \, \, ,
\end{eqnarray}
where $\bs{q}\,(\bs{k}\,)$ is the photon (pion) momentum, and $\pi=0,\pm 1$
is the pion charge in the intermediate state. The total pion-nuclear energy is
denoted by $E(k) = E_\pi (k) + E_A(k)$, the reduced mass is given by
$M(k) = E_\pi(k) E_A(k)/E(k)$ and $a=(A-1)/A = 2/3$.  In the
framework of the impulse approximation $V_{\pi \gamma}$ is expressed in terms
of the free pion-nucleon photoproduction t-matrix:
\begin{eqnarray} \label{cc2}
V_{\pi \gamma}(\bs{k},\bs{q})=
-\frac{\sqrt{M(q)M(k)}}{2\pi}<\pi(\bs{k}),f \mid
\sum^A_{j=1} \hat{t}_{\gamma N}(j)\mid i,\gamma (\bs{q})> ,
\end{eqnarray}
where $\mid i>$ and $\mid f>$ denote the nuclear initial and final
states, respectively, and  $j$ refers to the individual target nucleons.
Detailed information about the way of construction of
$V_{\pi \gamma}(\bs{k},\bs{q})$ is given in our previous paper \cite{
Kam91}.
Note only that for $\hat{\rm t}_{\gamma N}$ we shall use the modern
unitary version of the Blomqvist-Laget amplitude \cite{Lag88},
which describes
the real and imaginary parts not only for the resonance magnetic $M_{1+}$
but also for the resonance $E_{1+}$ multipole.

   Using the KMT version of multiple scattering theory \cite{Ker59} the pion
scattering amplitude $F_{\pi' \pi}$ is constructed as a solution of the
Lippmann-Schwinger equation
\begin{eqnarray} \label{cc3}
F_{\pi' \pi}(\bs{k}\,',\bs{k}) =V_{\pi' \pi}(\bs{k}\,',\bs{k})-
\frac{a}{(2\pi)^2} \sum_{\pi''} \int \frac{d^3k''}{M(k'')}
\frac{V_{\pi'\pi''}(\bs{k}\,',\bs{k}\,'')F_{\pi''\pi}(\bs{k}\,'',
\bs{k})}{E(k) - E(k'') + i\epsilon} \, \, .
\end{eqnarray}
Here  the pion-nuclear interaction is described by the first-order potential
$V_{\pi'\pi}$ which is related to the free $\pi N$ scattering
t-matrix \cite{Gmi85}

\begin{figure}[htbp]
\centerline{\psfig{figure=ali20.ps,height=6.2cm}}

\vspace{0.2cm}
{\small {\bf Fig. 20:}
Differential cross section for $^3He(\gamma,\pi^+)^3H$.
{\bf (a)} Angular distribution at $E_\gamma$=300 MeV
calculated with the full three-body wave function \cite{Bra75} and
the full production operator \cite{Lag88}.  Solid, dash-dotted and
dashed curves are the
results obtained in the coupled-channels approach that includes two-step
processes, in DWIA and PWIA, respectively. Experimental data are from
ref \cite{Hos88}($\circ$) and ref \cite{Bac73}($\bullet$).
{\bf (b)} Energy distribution at $\theta_{cm}=90^\circ$ calculated in the
coupled-channels framework. The solid (dash-dotted) lines are calculations
with (without) $E_{1+}(\Delta)$, the dashed curve is without the D-state
components in $^3$He.
The data points are from ref. \cite{Bel87}. }
\end{figure}

For eta photoproduction the formalism of eqs.~(\ref{cc1}-\ref{cc3}) can
easily be adopted. Asymptotic pionicstates $\pi$ are replaced by $\eta$ and
intermediate states $\pi'$ and $\pi''$ have to be summed over both $\pi$ and
$\eta$ states. In the case of pion production this is not necessary as long
as the pion energy is below the $\eta$ threshold. In fig. 20 we show the
result of our coupled-channels calculation in comparison with experimental
data. Fig. 20a compares the plane wave approximation (PWIA) with the
distorted wave impulse approximation (DWIA) that includes only $\pi^+$
rescattering and also with the full coupled channels calculation including
furthermore the very important single charge exchange (SCE) contribution
from intermediate $\pi^0$. Fig. 20b demonstrates the rather small
sensitivity to the different $\Delta$-resonance models with and without the
quadrupole $E2$ excitation. The agreement with the experimental data is
very good in both cases.

It is not surprising that the $E2(\Delta)$ is rather invisible, the same is
known for the nucleon. In order to study this effect, polarization
observables as the photon asymmetry $\Sigma$ are much more sensitive.
In a very simple model using harmonic oscillator
S-shell nuclear wave functions and excluding pion rescattering.
In this case we obtain simple relations between the
observables for $^3$He$(\gamma, \pi^+)^3$H and the elementary process
$p(\gamma, \pi^+)n$
\begin{eqnarray} \label{cc6}
\Sigma(^3{\mrm{He}}) =\Sigma (p),\qquad
T(^3{\mrm{He}}) = -P(p)\quad {\mrm {and}}\qquad
P(^3{\mrm{He}}) = -T(p),
\end{eqnarray}
provided the Lorentz transformation from the $\pi N$ $cm$ to the
$\pi^3$He $cm$ system has been taken into account.
The minus sign in eq. (\ref{cc6}) is due to an opposite sign
between spin-flip and non spin-flip matrix elements in $^3$He($\gamma,
\pi^+)^3$H compared to the free process $p(\gamma, \pi^+)n$.  This is a result
of the Pauli principle forbidding non spin-flip transitions on that proton in
$^3$He whose spin is aligned with that of the neutron.

\begin{figure}[htbp]
\centerline{\psfig{figure=ali21.ps,height=6.3cm}}

\vspace{0.2cm}
{\small {\bf Fig. 21:}
Comparison of the polarized photon asymmetry $\Sigma$ for $\gamma,\pi^+$
on proton and $^3He$ at $\theta_{cm}=90^\circ$ as a function of the
photon lab energy. {\bf (a)} Dashed and dotted lines are with and without
the $E_{1+}(\Delta)$ multipole \cite{Lag88}. {\bf (b)} Dashed and dotted lines
are
same as in (a) for a simple S-shell three-body wave function. The solid
and dash-dotted lines are with and without the $E_{1+}(\Delta)$, however,
additionally with the D-state components of the realistic three-body
wave function.}
\end{figure}

In fig. 21 we compare the photon asymmetry between proton and $^3He$. Apart
from the Lorentz transformation effect the simple relation of (\ref{cc6})
is observed. However, this changes drastically for a realistic $^3He$
calculation. In realistic models, wave functions are calculated as
solutions of the Faddeev equations with NN potentials, leading to about
8\% D-state components in $^3He$ \cite{Bra75}. From these D-states we obtain an
enhancement of the $E2(\Delta)$ due to an interference with the large
Kroll-Ruderman $E_{0+}$ multipole, which is absent for nucleons in the
S-shell. Explicitly, for $\theta_{cm}=90^\circ$, the contribution
additional to the photon asymmetry $\Sigma(p)$ on the proton has the
following form
\begin{eqnarray} \label{cc11}
\Sigma_{SD}(90^{0}) \sim {\mrm S}_{\mrm M}(Q)\,{\mrm D}_{\mrm M}(Q)
\left[\,{\mrm{Re}}E_{0+}f_{EM}^{*}-\frac{q}{2k}\mid E_{0+}\mid^2-
\frac{4q^2+k^2}{2kq}\mid f_{EM}\mid^2\right]\,
\end{eqnarray}
with $f_{EM}=3E_{1+}-M_{1+}+M_{1-}$ and the S- and D-wave spin-flip
formfactors $S_M(Q)$ and $D_M(Q)$ for $^3He$.

\begin{figure}[htbp]
\centerline{\psfig{figure=ali22.ps,height=11cm}}

\vspace{0.2cm}
{\small {\bf Fig. 22:}
Differential cross section for eta photoproduction on $p$, $n$, $d$,
$^3$He, $^3$H and $^4$He in plane wave approximation (PWIA).}
\end{figure}

Finally in fig. 22 we show the differential cross sections for all light nuclei
up to $^4$He. While the angular distribution is rather flat for nucleons,
as expected from s-wave dominance, it appears more and more peaked in forward
direction for $A>1$. This reflects the signature of the nuclear formfactors
as the momentum transfer in $\eta$ photoproduction is rather large,
$Q^2$ = 7.8 $fm^{-2}$ at threshold. The biggest cross section can be expected
for the trinucleon; it is proportional to the free nucleon cross section
multiplied by the square of the trinucleon formfactor.
However around $90^\circ$ the cross section on the deuteron
gains over the trinucleon, in particular due to the not yet understood
enhancement seen in the experiment \cite{And69}. The coherent cross section
for $^4He$
vanishes for $\Theta=0$ and reaches roughly the $10 nb$ level in a small
angular region. For most angles it falls below $1 nb$.

\begin{center}
{\bf 9. Summary and conclusions}
\end{center}

Investigations with electromagnetic interactions have contributed
substantially to our knowledge of the structure of hadrons. With the
new electron accelerators, polarization degrees of freedom will play a
decisive role in unravelling the unsettled questions. Typical examples
range from the distribution of the neutron's charge to the spin content
and the strange content of the nucleon.

Photo- and electroproduction of mesons from the nucleons promise to
determine some of the most wanted quantities at intermediate energies,
e.g. the $L_{1+}$ and $E_{1+}$ multipole of the $\Delta$ excitation,
the $L_{1-}$ and $M_{1-}$ of the Roper resonance, the high rate for
$\eta$ production near the $S_{11}(1535)$ and the very low $\eta$ rate
for the second $S_{11}(1650)$. Also the production of two pions or more
in order to test consequences of chiral symmetry and the interaction of
very low energy pions. The strange behaviour of the threshold $E_{0+}$
amplitude for $\gamma, \pi^0$ on th eproton has added another challenge
to future experiments. New data has been obtained and is beeing analysed
that connect the threshold region to the well established first resonance
region. Polarized observables will be measured in a next generation of
experiments, allowing also precise determinations of small p-wave
multipoles at threshold as well as the phases of the $E_{0+}$ amplitude.
These phases that are connected to very low pion nucleon scattering
will be a rather stringent test of models for the $\pi N$ system as
for low energy QCD.

With the completion of the new high duty-factor electron accelerators at
Mainz and Bonn, the eta production has become very attractive. New data
that is analysed in preliminary form show an amazing accuracy in energy
resolution and in absolute quantity. A first result of the data is the
determination of the $\eta NN$ coupling, long in discussion in various
experiments and theoretical models. Comparing to the angular distributions
of high quality only a pseudoscalar coupling with a rather small coupling
constant of $g_{\eta NN}^2/4\pi$ around 0.2 -- 0.5 is possible.
Recently, Piekarewicz \cite{Pie93} has estimated the $g_{\eta NN}$ coupling
constant in a rather indirect way from the $\pi-\eta$ mixing amplitude in
the hadronic model where the mixing was generated by $\bar N N$ loops and
thus driven by the proton-neutron mass difference. To be in agreement with
results from chiral perturbation theory the $\eta N$ coupling had to be
constained to the range of 0.32 - 0.53.
Pseudovector
models give quite different results in photoproduction due to the anomalous
magnetic moments of the nucleon. It is this fact, why, in contrast to
hadronic reactions, it is easy to rule it out. Accidently it turns out that
the background contribution for eta production is almost vanishing, due to
the cancellation of vector meson exchange and pseudoscalar Born terms.
This fact will provide a veryu accurate determination of resonance
parameters for the $S_{11}(1535)$. Due to the nearby threshold, this is not
uniquely possible in $\pi N$ scattering and teh available $\pi,\eta$ data
are very inaccurate.

The next series of experiments on meson production will attack more
difficult issues as two-pion production, eta production on light nuclei,
quasifree and coherent, and pion and eta production on polarized targets
using polarized photon beams. The large number of structure functions in
the coincidence cross section and of the polarization observables in
photoproduction will allow to disentangle nucleon resonances as the
$P_{11}(1440)$ or $D_{13}(1520)$ that are normally hidden in the
unpolarized cross sections. All of these experiments will require a high
degree of precision and a careful analysis of the systematic errors.
However, they will help to increase our knowledge on the structure of
hadrons and their underlying symmetries.

\begin{center}
{\bf References}
\end{center}

\vspace{-2cm}

\renewcommand{\refname}{}

\end{document}